\documentclass{emulateapj}
\usepackage{subfig}
\usepackage{multirow}
\def\msun{$M_{\odot}$}

\shortauthors{Lin et al.}
\begin{document}

\title{The Megasecond {\it Chandra} X-Ray Visionary Project
  Observation of NGC 3115 (II): properties of point sources}

\author{Dacheng Lin\altaffilmark{1,2}, Jimmy A. Irwin\altaffilmark{2},
  Ka-wah Wong\altaffilmark{3}, Zachary G. Jennings\altaffilmark{4}, Jeroen Homan\altaffilmark{5}, Aaron
  J. Romanowsky\altaffilmark{4,6}, Jay Strader\altaffilmark{7},
  Gregory R. Sivakoff\altaffilmark{8}, Jean P. Brodie\altaffilmark{4},
  Ronald A. Remillard\altaffilmark{5}}

\altaffiltext{1}{Space Science Center, University of New Hampshire, Durham, NH 03824, USA, email: dacheng.lin@unh.edu}
\altaffiltext{2}{Department of Physics and Astronomy, University of
  Alabama, Box 870324, Tuscaloosa, AL 35487, USA}
\altaffiltext{3}{Eureka Scientific, Inc., 2452 Delmer Street Suite 100, Oakland, CA 94602-3017}
\altaffiltext{4}{University of California Observatories, Santa Cruz, CA 95064, USA}
\altaffiltext{5}{MIT Kavli Institute for Astrophysics and Space Research, MIT, 70 Vassar Street, Cambridge, MA 02139-4307, USA}
\altaffiltext{6}{Department of Physics and Astronomy, San Jos\'{e} State University, One
Washington Square, San Jos\'{e}, CA 95192, USA}
\altaffiltext{7}{Department of Physics and Astronomy, Michigan State University,
East Lansing, Michigan, MI 48824, USA}
\altaffiltext{8}{Department of Physics, University of Alberta, Edmonton, Alberta,
T6G 2E1, Canada}

\begin{abstract}
  We have carried out an in-depth study of low-mass X-ray binaries  (LMXBs) detected in the nearby lenticular galaxy NGC 3115, using the  Megasecond \textit{Chandra} X-Ray Visionary Project observation  (total exposure time 1.1 Ms). In total we found 136 candidate LMXBs  in the field and 49 in globular clusters (GCs) above 2$\sigma$  detection, with 0.3--8 keV luminosity $L_\mathrm{X}$  $\sim$10$^{36}$--10$^{39}$ erg~s$^{-1}$. Other than 13 transient  candidates, the sources overall have less long-term variability at  higher luminosity, at least at $L_\mathrm{X}\gtrsim2\times10^{37}$  erg~s$^{-1}$. In order to identify the nature and spectral state of  our sources, we compared their collective spectral properties based  on single-component models (a simple power law or a multicolor disk)  with the spectral evolution seen in representative Galactic  LMXBs. We found that in the $L_\mathrm{X}$ versus photon index  $\Gamma_\mathrm{PL}$ and $L_\mathrm{X}$ versus disk temperature  $kT_\mathrm{MCD}$ plots, most of our sources fall on a narrow track  in which the spectral shape hardens with increasing luminosity below  $L_\mathrm{X}\sim7\times10^{37}$ erg~s$^{-1}$ but is relatively  constant ($\Gamma_\mathrm{PL}$$\sim$1.5 or  $kT_\mathrm{MCD}$$\sim$1.5 keV) above this luminosity, similar to  the spectral evolution of Galactic neutron star (NS) LMXBs in the  soft state in the \textit{Chandra} bandpass. Therefore we identified  the track as the NS LMXB soft-state track and suggested sources with  $L_\mathrm{X}\lesssim7\times10^{37}$ erg~s$^{-1}$ as atolls in the  soft state and those with $L_\mathrm{X}\gtrsim7\times10^{37}$  erg~s$^{-1}$ as Z sources. Ten other sources (five are transients)  displayed significantly softer spectra and are probably black hole  X-ray binaries in the thermal state. One of them (persistent)  is in a metal-poor GC.
\end{abstract}


\section{INTRODUCTION}
\label{sec:intro}
\begin{figure} 
\centering
\includegraphics[width=3.4in]{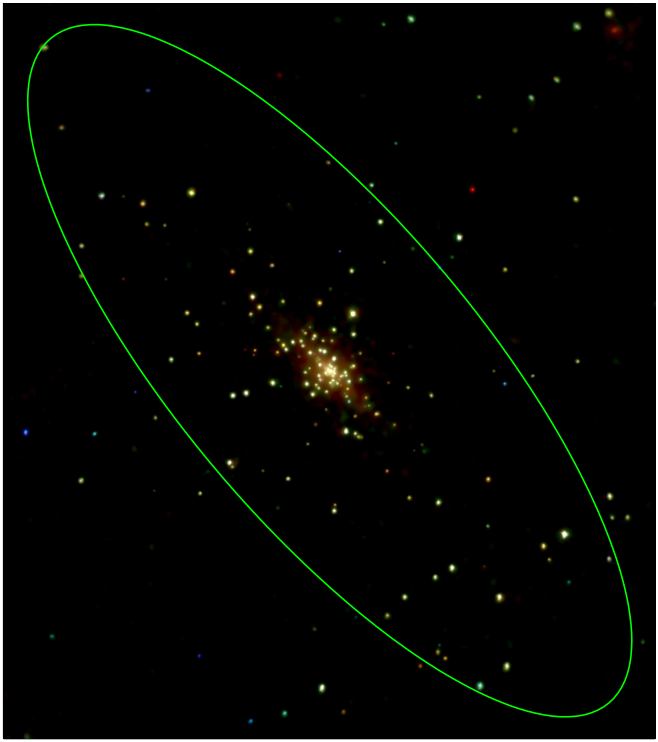}
\caption{{\it Chandra} X-ray image of NGC 3115. The image is
  false-colored using adaptively smoothed (with the CIAO task
  csmooth), exposure corrected images in 0.5--1.2 keV (red), 1.2--2
  keV (green), and 2--7 keV (blue). The $D_\mathrm{25}$ ellipse of the
  galaxy is also shown. The diffuse emission near the galaxy center
  comes from diffuse hot gas, unresolved LMXBs, and other unresolved
  stellar emission \citep{woiryu2011,woirsh2014}.
  \label{fig:colorimage}}
\end{figure}

In a low-mass X-ray binary (LMXB), a neutron star (NS) or a
stellar-mass black hole (BH) accretes matter from a Roche-lobe
filling, low-mass companion star via an accretion disk. Copious X-rays
are produced in the inner disk \citep{shsy1973}. In the case of NS
LMXBs, there is also strong X-ray emission from the boundary layer
formed by the settling of the accretion flow onto the NS surface
\citep{insu1999,posu2001}.  Our knowledge of LMXBs was revolutionized
thanks to 16 years of intensive observations of such objects in our
Galaxy by the {\it Rossi X-ray Timing Explorer} \citep[{\it
  RXTE},][]{brrosw1993}.

BH LMXBs constitute the majority of BH X-ray binaries (BHBs) known in
our Galaxy, though a few BHBs with a high-mass stellar companion also
exist. BHBs exhibit three main X-ray spectral states: the hard state,
the thermal state, and the steep power law (SPL) state
\citep{remc2006,mcre2006}. These states are normally described with
spectral models consisting of two main continuum components: a
standard thermal multi-color disk \citep[MCD, {\it diskbb} in
XSPEC,][]{ar1996} and a Comptonized component (often modeled with a
single power law, PL, or the Comptonization model {\it comptt} in
XSPEC). The hard state, which tends to occur below several percent of
the Eddington luminosity ($L_\mathrm{Edd}$), is characterized by a
strong Comptonized component dominating the spectra at least above
$\sim$2 keV and extending to more than 100 keV. A weak, cool thermal
disk might also be detected, with the maximum disk temperature
$kT_{\rm MCD}$$\ll$1 keV. The thermal state, which is normally
observed at high luminosities, is characterized by a strong disk
component with $kT_\mathrm{MCD}\sim$ 0.5--1 keV and the disk fraction
of $>75\%$ in 2--10 keV \citep[as defined in][]{remc2006}. A weak
Comptonized component is also often present but only dominates at high
energies. The SPL state tends to occur at very high luminosities
($\gtrsim$0.1 $L_\mathrm{Edd}$), and the hallmark of this state is a
strong Comptonizaton component with photon index
$\Gamma_\mathrm{PL}\sim 2.5$ \citep{remc2006}. A sizable thermal
component is also normally seen in this state. Almost all known Galactic BH
LMXBs are transients. They often show hysteresis in transitions to and
from the hard state, with the transition to the hard state occuring at
luminosities tightly clustered around 2\% of the Eddington value
\citep{ma2003}, and the transition from the hard state at higher but
more scattered luminosities
\citep{mikiha1995,maco2003a,hobe2005}. Similar hysteresis behavior
is observed in transient NS LMXBs.

There are two main classes of NS LMXBs, atoll and Z sources
\citep{hava1989,va2006}, named after the patterns that they trace out
in X-ray color-color diagrams (CDs) or hardness-intensity diagrams
(HIDs).  Atolls radiate at $\sim$0.001--0.5 $L_\mathrm{Edd}$ and trace
out their patterns in CDs/HIDs on timescales of weeks to months. They
have two main distinct spectral states, i.e., hard
($\Gamma_\mathrm{PL}\lesssim$2, extending to 100 keV or above) and
soft states (most emission $\lesssim$20 keV). The hard state tends to
be observed at low luminosity ($\lesssim$0.1 $L_\mathrm{Edd}$), while
the soft state is normally observed at high luminosity (larger than a
few percent $L_\mathrm{Edd}$). Spectra in the ``transitional'' state
between these two are also occasionally seen. Z sources are more
luminous than atolls, at near or above Eddington luminosity, and they
trace out roughly Z-shaped tracks in CDs/HIDs within a few days (i.e.,
faster than atolls), with X-ray spectra that are generally
soft. Thanks to a recent transient Z source \object{XTE J1701-462},
which exhibited the Z-source characteristics when it was accreting at
near or above Eddington luminosity and transitioned to an atoll during
the decay of its 2006-2007 outburst, we now know that Z and atoll
sources are essentially the same type of objects at different mass
accretion rates \citep{lireho2009,hovafr2010}. Unlike BHBs, whose
  tracks in the CDs/HIDs tend to show large scatter, mainly due to the
  presence of the SPL state, NS LMXBs tend to trace out clear narrow
  tracks in the CDs/HIDs, which depend mostly on the accretion rate
  \citep{dogi2003,remc2006,frhore2015}. 

The spectral modeling for NS LMXBs is complicated by the presence of
the boundary layer emission and is relatively controversial, compared
to BHBs. \citet{lireho2007} used a similar spectral model to the one
used for BHBs (described above) except for an additional
single-temperature blackbody (BB) to describe the boundary layer and
were able to infer $L\propto T^4$ trends for both the disk and the
boundary layer in the soft state of two atolls. Such a trend for the
disk is often observed for the thermal state of BHBs and is expected
if the disk is essentially thermal and is truncated at the innermost
stable circular orbit (ISCO). \citet{lireho2009} applied this model to
\object{XTE J1701-462} and also observed the $L\propto T^4$ trends for
both the disk and the boundary layer in its atoll stage. However, such
trends were not observed in its Z-source stage, because the inner disk
and the boundary layer both reach the local Eddington limit in this
stage so that the increase in the accretion rate tends to lead to an
increase in the emission area with relatively constant temperature
(increasing the inner disk radius at constant inner disk temperature
for the accretion disk).

Our knowledge of X-ray binaries in nearby galaxies has significantly
increased since the launch of {\it XMM-Newton} and {\it Chandra X-ray
  Observatory} in 1999. In particular, {\it Chandra}'s superb spatial
resolution and excellent sensitivity \citep{wibrca2002} allow for
population studies of X-ray binaries in a single galaxy with
snapshots, which is important for understanding the origin and
evolution of such sources. To fully understand these sources, one key
task is to identify their nature (i.e., BH versus NS X-ray binaries)
and the X-ray spectral state. To achieve this, early studies
  tried to stack sources in luminosity ranges in order to improve the
  statistics \citep[][]{makuze2003,iratbr2003}. Recently, relatively detailed
studies of individual sources in nearby galaxies were carried out
\citep[e.g.,][]{brfabl2010,bafaze2012,burakr2013,bagamu2013,bagamu2014},
but limited by statistics, these studies have been mostly focused on
the few most luminous sources (above several $10^{37}$ erg s$^{-1}$)
in each galaxy. It turns out that the differentiation between BH and
NS X-ray binaries is generally very difficult. This is because most
sources have X-ray spectra that are relatively hard
($\Gamma_\mathrm{PL}<2.5$) in the {\it Chandra} and {\it XMM-Newton}
bandpass (about 0.3--8 keV) and it is difficult to determine whether
they are BHBs in the hard state or NS LMXBs in the soft state (they
are unlikely to be NS LMXBs in the hard state due to the high
luminosities of the sources studied). NS LMXBs in the soft state can
appear hard in such an energy band because of the presence of the hot
boundary layer component in the X-ray spectra
\citep[e.g.,][]{lireho2010,lireho2012}.

\object{NGC 3115} was selected as the target of a 1 Megasecond {\it
  Chandra} X-ray Visionary Project (XVP) in Cycle 13. The goals were
to study the gas flow inside the Bondi radius of the central
supermassive black hole and obtain a deep look at the X-ray binary
population of a normal early-type galaxy. The former has been
presented in \citet{woirsh2014}. For the X-ray binaries, we have
presented the X-ray luminosity function in Lin et al. (2014, Paper I
hereafter), and here we concentrate on the detailed properties of
discrete sources, such as the long-term spectral and flux variability
and the spectral characteristics. One main goal of our study is to
identify their nature, but different from previous studies, we will
achieve this by systematic comparison of our sources at various
luminosity levels with Galactic X-ray binaries. \object{NGC 3115} is a
lenticular (S0) galaxy with an age of $\sim$8.4 Gyr \citep{sagoca2006}
and at a distance of 9.7 Mpc \citep{todrbl2001}. Including previous
observations, the total exposure time of {\it Chandra} on this galaxy
is $\sim$1.1 Ms, and the limiting X-ray luminosity is $\sim$$10^{36}$
erg s$^{-1}$, making it one of {\it Chandra}'s best observed normal
early-type galaxies.

In Section~\ref{sec:reduction}, we describe the source detection, the
calculation of flux, the simple spectral fits, the measurement of
long-term and short-term variability, and multiwavelength
cross-correlation. In Section~\ref{sec:res}, we present the various
properties of X-ray binaries in NGC 3115, including the long-term
variability and spectral characteristics that are used to classify the
sources. We further discuss the possible nature of our sources in
Section~\ref{sec:dis}.  We present our conclusions in
Section~\ref{sec:con}.

\section{DATA ANALYSIS}
\label{sec:reduction}

\subsection{Observations and Source Detection}

\tabletypesize{\scriptsize}
\setlength{\tabcolsep}{0.02in}
\begin{deluxetable}{rrccc}
\tablecaption{Observation Log\label{tbl:obslog}}
\tablewidth{0pt}
\tablehead{\colhead{Notation} & \colhead{Obs. ID} &\colhead{Date} &\colhead{Exposure} &\colhead{Offset\tablenotemark{a}}\\
& &  & (ks) & (arcmin)}
\startdata
1 & 2040 & 2001-06-14 & 35.8 & 1.5\\
2 & 11268 & 2010-01-27 & 40.6 & 0.1\\
3 & 12095 & 2010-01-29 & 75.6 & 0.1\\
4 & 13817 & 2012-01-18 & 171.9 & 0.0\\
5 & 13822 & 2012-01-21 & 156.6 & 0.0\\
6 & 13819 & 2012-01-26 & 72.9 & 0.0\\
7 & 13820 & 2012-01-31 & 184.1 & 0.0\\
8 & 13821 & 2012-02-03 & 157.9 & 0.0\\
9 & 14383 & 2012-04-04 & 119.4 & 0.3\\
10 & 14419 & 2012-04-05 & 46.3 & 0.3\\
11 & 14384 & 2012-04-06 & 69.7 & 0.3
\enddata 
\tablenotetext{a}{Aim point offset from observation 13820.}
\end{deluxetable}

The eleven {\it Chandra} observations of NGC 3115 are listed in
Table~\ref{tbl:obslog}. They were made during three epochs: one in 2001,
two in 2010, and nine in 2012. We hereafter refer to them as Obs
1--11 in chronological order (Table~\ref{tbl:obslog}).
All observations used the imaging array of the AXAF CCD Imaging
Spectrometer \citep[ACIS; ][]{bapiba1998}. We analyzed the data with
the {\it Chandra} Interactive Analysis of Observations (CIAO, version
4.6) package.  We reprocessed the data to apply the latest calibration
(CALDB 4.5.9) and the subpixel algorithm \citep{likapr2004} using the
CIAO script {\it chandra\_repro}. Background flares are only clearly
seen in Obs 1, 5, and 6, for only very short durations. We excluded
them if they are higher than 4$\sigma$ above the mean background
level. In this way, we excluded 1.2 ks, 3.6 ks, and 2.7 ks data for
Obs 1, 5, and 6, respectively. The final exposure used for each
observation is given in Table~\ref{tbl:obslog}.

To increase the detection sensitivity, we combined all 11 observations
to create a deep merged observation (Obs {\it Esum} hereafter) using
the CIAO script {\it merge\_obs}. To correct for relative astrometry
between different observations, we created new aspect solution files
by comparing the source list from each single observation to the
source list from a single reference observation, which we chose to be
the longest observation (13820). We only used bright sources
($>$$6\sigma$) with off-axis angles $<$$6\arcmin$ in the
cross-correlation. The average separation residuals of the source
matches are 0.1$\arcsec$ after relative astrometry correction. The new
aspect solution files were then applied to the event files and
subsequent analysis.

We used the CIAO {\it wavdetect} wavelet-based source detection
algorithm \citep{frkaro2002} to search for discrete X-ray sources. The
search was done twice, first over the single observations to determine
the relative astrometry correction described above and the second time
over the (relative astrometry corrected) single observations and the
merged one. The count images were made in the broad ($b$) energy band
0.5--7.0 keV adopted in the {\it Chandra} Source Catalog (CSC), while
the exposure maps were constructed at the corresponding monochromatic
effective energy, \citep[i.e., 2.3 keV,][]{evprgl2010}. The point
spread function (PSF) maps used correspond to the 50\% enclosed counts
fraction (ECF) at 2.3 keV. For the merged observation, the PSF map was
obtained by averaging those from single observations weighted by the
exposure. We used two different resolutions: one at single sky pixel
resolution (0\farcs492) over the full field of view (FOV) and the
other at 1$/$8 sky pixel resolution covering an area of
3$\arcmin$$\times$3$\arcmin$ centering at the center of NGC 3115. The
subpixel binning images were used to improve the spatial resolution in
the crowded field near the center of the galaxy. The limiting
significance level was set to $10^{-6}$, which formally corresponds to
$\sim$1 false source due to random fluctuations per 1024$\times$1024
image pixels (about one CCD area for images at single sky pixel
resolution). The wavelet scales were set to 1, 2, 4, 8, 16, and 32
image pixels for images at single sky pixel resolution and 4, 8, 16,
32, 64, 128, and 256 image pixels for images at 1/8 sky pixel
resolution.

Sources detected from the merged observation and single observations
were cross-correlated to create the unique source list. Starting from
the source list from the merged observation, we searched for new
sources detected in single observations but not in the merged
one. Such sources should be faint and highly variable or
transient. Sources are deemed to be duplicates if their separation is
less than their combined 99.73\% (i.e., 3$\sigma$) statistical
positional uncertainty or if their 50\% PSF circular regions overlap
with each other across different observations. The statistical
positional uncertainties that we used are based on Equation (12) of
\citet{kikiwi2007}, which provides the 95\% statistical positional
uncertainty as a function of the source net counts and the off-axis
angle based on a large number of simulations using the {\it Chandra}
simulation tool MARX. The 95\% positional uncertainties are converted
to the 99.73\% errors by multiplying by a factor of 1.405 (this
assumed a two-dimensional, circularly symmetric Gaussian distribution
for the source position from {\it wavdetect}). As a check on the above
matching criteria, we also tested a smaller searching radius by using
the 95\% statistical positional uncertainty only (the 50\% PSF
overlapping is not used). In the end 95 more sources were found. From
visual inspection, we found that six of them lie in the central
crowded field and two at the CCD edge, thus probably having large
systematic positional uncertainties and explaining their relatively
large offsets from detections from the merged observation. The
remaining sources generally have their separations from those of the merged
observation much less than the size of their region ellipse from {\it
  wavdetect}. Thus in the end we did not treat any of these 95 sources
as new sources and used the source list obtained above using the
99.73\% statistical positional uncertainty and the 50\% PSF circular
region.

\subsection{Flux and Spectral Characterization}
After obtaining a merged source list, we extracted the spectrum for
each source for each single observation. The source region was set to
be a circle enclosing 90\% of the PSF at 2.3 keV. The background
region was set to be a concentric annulus, with inner and outer radii
of two and five times the source radius, respectively. Nearby sources,
if present, were excluded from the source and background regions, but
the inner circular source region enclosing 50\% of the PSF was not
excluded. We used the CIAO task {\it mkacisrmf} to create the response
matrix files and the CIAO tasks {\it mkarf} and {\it arfcorr} to
create the point-source aperture corrected auxiliary response
files. The spectral and response files corresponding to the merged
observation were created using the CIAO task {\it combine\_spectra}.

The background-subtracted count rates (but not aperture-corrected) in
different energy bands were obtained from the spectral files. To
correctly determine confidence bounds for low count limits, we used
the CIAO task {\it aprates}. The conversion from the count rates to
the fluxes were based on the response files and assumed an absorbed PL
spectral shape with $\Gamma_\mathrm{PL}=1.7$ and Galactic absorption
$N_\mathrm{H}=4.32\times10^{20}$ cm$^{-2}$ \citep{kabuha2005}. Throughout
the paper, all fluxes and luminosities quoted (including those
obtained from spectral fits described below) are corrected for
Galactic absorption (but not intrinsic absorption, unless indicated
otherwise).

To characterize the spectral properties of our sources, we calculated
the hardness ratios ${\rm HR}=(H-S)/(H+S)$, where S and H are the
energy fluxes in the soft and hard energy bands, respectively, using
the method of Bayesian estimation \citep{pakasi2006}. We also carried
out simple spectral fits to spectra above 4$\sigma$ using two
single-component models: a PL and a MCD. Due to the low statistics of
most sources, we binned the spectra to have a minimum of one count per
bin and used the C statistic in the fits. Both models included
absorption (we used the {\it wabs} model in XSPEC; we found no
significant effect on our results if we chose other absorption models
such as {\it tbabs}, due to little absorption of most of our sources),
with the minimum set to be the Galactic value of $N_{\rm
  H}=4.32\times10^{20}$ cm$^{-2}$.

\subsection{Long-term and Short-term Variability}
\label{sec:calvar}
The variability of the source was measured in several aspects. We defined the
long-term flux variability as $V_\mathrm{var}=F_\mathrm{max}/F_\mathrm{min}$
and the significance of the difference as
\begin{equation}
S_\mathrm{var}=\frac{F_\mathrm{max}-F_\mathrm{min}}{(\sigma_\mathrm{max}^2+\sigma_\mathrm{min}^2)^{1/2}},\label{eq:svar}
\end{equation}
where $F_\mathrm{max}$ and $F_\mathrm{min}$ are the maximum and minimum
0.5--7.0 keV fluxes of a unique source among the single observations,
with the corresponding errors $\sigma_\mathrm{max}$ and $\sigma_{\rm
  min}$, respectively. We only used detections with the flux above
twice the error ($\sigma$) when calculating $F_\mathrm{max}$ (if no
detections above $2\sigma$, the one with the highest significance was
used as $F_\mathrm{max}$), while we used $2\sigma$ as the flux for
detections with the flux less than $2\sigma$ when calculating $F_{\rm
  min}$.

We measured the short-term variability using the Gregory-Loredo
algorithm \citep{grlo1992} implemented by the CIAO tool {\it glvary}
\citep{evprgl2010}. It splits the events into multiple time bins and
looks for significant deviations. The variation of the effective area
with time was taken into account and was obtained by another CIAO tool
{\it dither\_region}. The different degrees of confidence are indicated
by the parameter of ``variability index'', which spans values within
[0, 10] and is larger for variability of higher confidence
\citep{evprgl2010}. 

\subsection{Multiwavelength Cross-correlation}
\label{sec:mulwavcc}
Accompanying the {\it Chandra} XVP observation, a six pointing {\it
  Hubble Space Telescope} ({\it HST}) mosaic observation in the F475W
and F850LP filters (hereafter $g$ and $z$) using the Advanced Camera
for Surveys (ACS) was also acquired in the field of NGC 3115. The
total field of view of this mosaic observation is slightly larger than
the $D_{25}$ region of NGC 3115, which has a semi-major axis of
$a=3.62\arcmin$ (10.2 kpc), a semi-minor axis of $b=1.23\arcmin$ (3.5
kpc) and a position angle of $40\degr$ \citep{dedeco1991}. The galaxy
was also imaged in $g$, $r$, $i$-band filters on 2008 January 4th
using Suprime-Cam on the 8.2-m Subaru telescope. In Paper I, we have
cross-correlated our X-ray sources with the 360 globular clusters
(GCs) from the {\it HST}/ACS mosaic imaging and the 421 ones from the
Subaru/Suprime-Cam imaging \citep{arrobr2011,jestro2014}. The match
was identified if the separation is less than the 99.73\% positional
uncertainty (combining the X-ray and optical components). The maximum
separation allowed is $2\arcsec$ in order to limit the spurious
rate. We note that before the cross-correlation, the systematic offset
between different source lists has been corrected through multiple
steps: the Subaru/Suprime-Cam astrometry was registered to the
USNO-B1.0 Catalog \citep{moleca2003} first, then the {\it HST}/ACS
astrometry was registered to the Subaru/Suprime-Cam one, and in the
end the astrometry of our X-ray sources was registered to the {\it
  HST}/ACS one (therefore the absolute astrometry of our X-ray sources
has been corrected). 

Table~\ref{tbl:gclmxb} lists the 37 matches with {\it HST}/ACS GCs
\citep[23 have the $g-z$ color $>1.13$ and thus are red/metal-rich,
while the other 14 have $g-z<1.13$ and are blue/metal-poor, following
the division in][]{jestro2014} and the 7 matches with
Subaru/Suprime-Cam GCs identified in Paper I. In Paper I, we also
identified five other sources whose optical counterparts were not
classified as GCs by \citet{jestro2014} but were assumed to be GC
candidates by us (Table~\ref{tbl:gclmxb}). Four of them (S12, S53,
S65, and S79) are within 0.25$D_{25}$, thus very unlikely to be AGNs
(Paper I). The other one (S92), at an outer region, has an optical
counterpart with the size and the color typical of GCs, though it has
a radial velocity from the spectroscopic measurement (238 km s$^{-1}$,
Table~\ref{tbl:gclmxb}) lower than typical values seen in other GCs
($>350$ km s$^{-1}$).

Adopting the same matching criteria, we also searched for the non-GC
counterparts to our X-ray sources outside 0.25$D_{25}$ from these
optical observations. Such matches are most probably cosmic X-ray
background sources (CXBs), especially active galactic nuclei (AGNs),
instead of the optical counterparts to field LMXBs. The
  optical emission of field LMXBs can achieve the maximum when the
  accretion rate is near the Eddington limit and the companion is an
  evolved star (so bright optical emission from both a large disk and
  a large companion), as seen in the Galactic Z source \object{Cyg X-2}
  \citep{vamc1995}. However, such sources are still below the
  detection limit of our optical images, by $\sim$3 mag. By rotating
the X-ray source positions around the center of the galaxy by
$\pm10\degr$, $\pm170\degr$, and $180\degr$ and using X-ray sources
above $4\sigma$, we estimated the rate of spurious matches to be about
3\% and 5\% for the {\it HST}/ACS counterparts and Subaru/Suprime-Cam
matches, respectively.

From visual inspection, we found that some very faint sources in the
{\it HST}/ACS observation were not detected by the tool SExtractor
used by \citet{jestro2014}.  Concentrating on the region within
(0.25--1)$D_{25}$, we visually identified five X-ray sources
coincident with such faint sources and assume them to be AGNs. Within
0.25$D_{25}$, we also visually found one source, i.e., S65, with a
faint counterpart not detected by \citet{jestro2014}, and we have
assumed it to be a GC, as mentioned above.

\section{RESULTS AND DISCUSSION}
\label{sec:res}
\begin{figure} 
\centering
\includegraphics[width=3.4in]{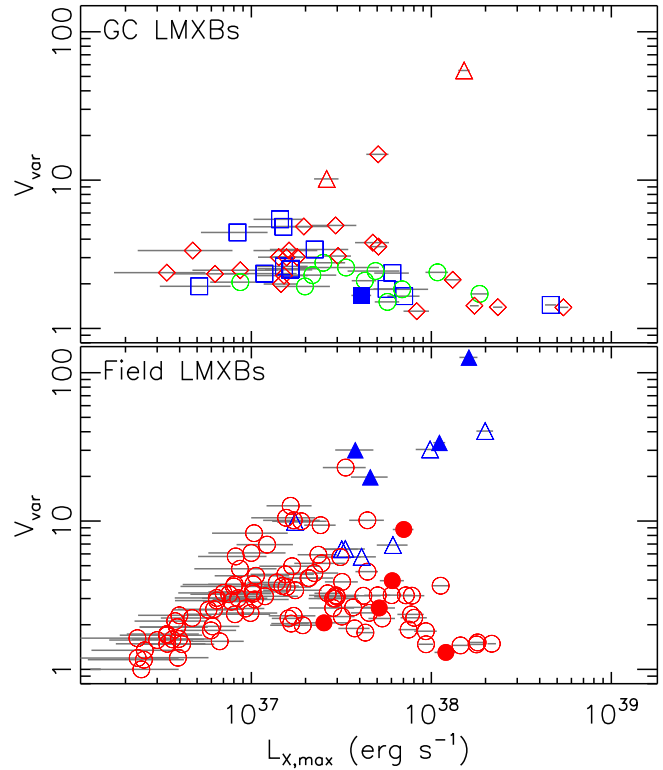}
\caption{The long-term variability versus the maximum 0.5--7 keV
  luminosity for all LMXBs excluding those in the central
  $a=10\arcsec$ ellipse. The top panel is for GC LMXBs, with the blue
  squares and red diamonds for the {\it HST}/ACS blue/metal-poor and
  red/metal-rich GCs from \citet{jestro2014}, respectively, and the
  green circles for other GCs (the Subaru/Suprime-Cam GCs and the
  extra five {\it HST}/ACS GC candidates identified by us;
  Section~\ref{sec:mulwavcc}). Two transients, both in red GCs, are
  plotted with red triangles. The bottom panel is for field LMXBs. The
  blue triangles denote the transient candidates. In both panels, the
  filled symbols denote BHCs (see text for
  details). \label{fig:ltvar_maxmin}}
\end{figure}

\begin{figure*}[!htpb]
  \centering
  \subfloat[Transient candidates with the outburst detected in the first
   epoch.]{%
    \includegraphics[width=0.8\textwidth]{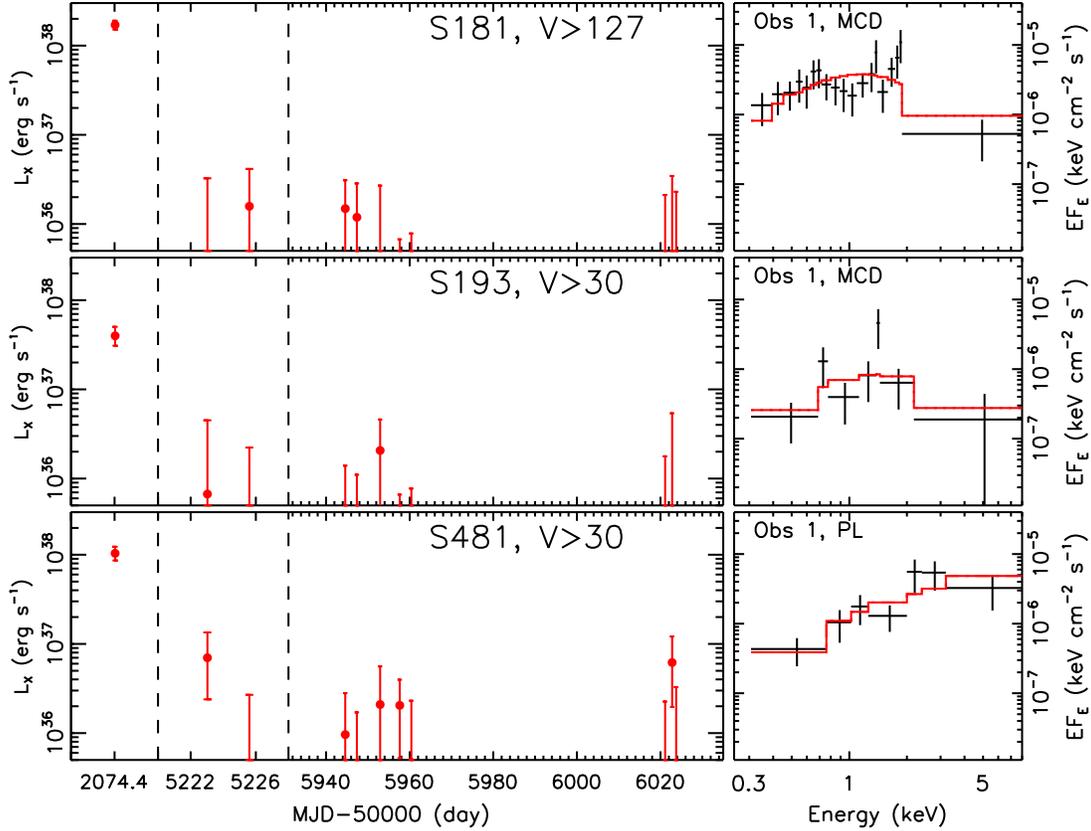}%
    \label{fig:lc_sub_epoch1}
  }\\
  \subfloat[Transient candidates with the outburst detected in the second
   epoch.]{%
    \includegraphics[width=0.8\textwidth]{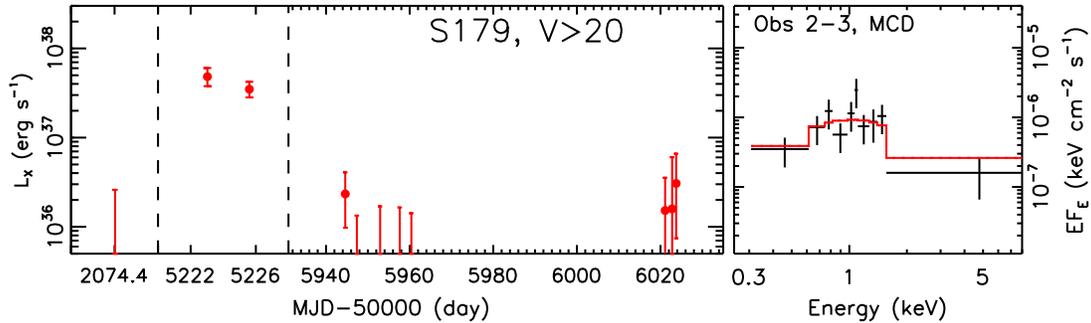}%
    \label{fig:lc_sub_epoch2}
  }
  \caption{Long-term light curve (left panels, with the source number
    and long-term variability $V$ (i.e., $V_\mathrm{var}$ in
    Section~\ref{sec:calvar})) and sample spectra (right panels, with
    annotations for the observations and spectral models used) for
    special sources.}
\end{figure*}

\begin{figure*}
  \centering
  \ContinuedFloat
  \subfloat[ Transient candidates with the outburst detected in the third
   epoch.]{%
    \includegraphics[width=0.8\textwidth]{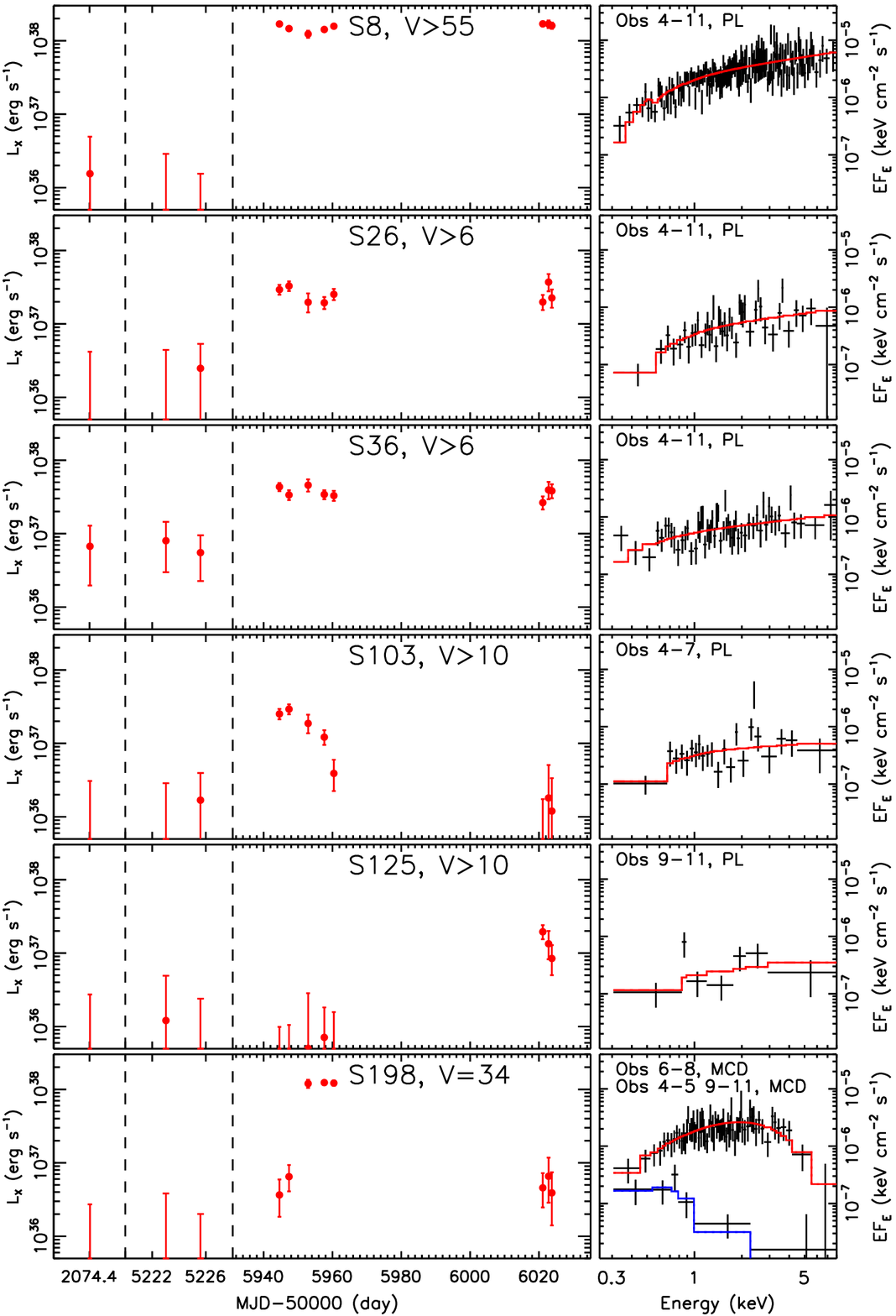}%
    \label{fig:lc_sub_epoch3}
  }
  \caption{(continue)}
\end{figure*}

\begin{figure*}
  \centering
  \ContinuedFloat
  \subfloat[Transient candidates with the outburst detected in the
  second and third epochs.]{%
    \includegraphics[width=0.8\textwidth]{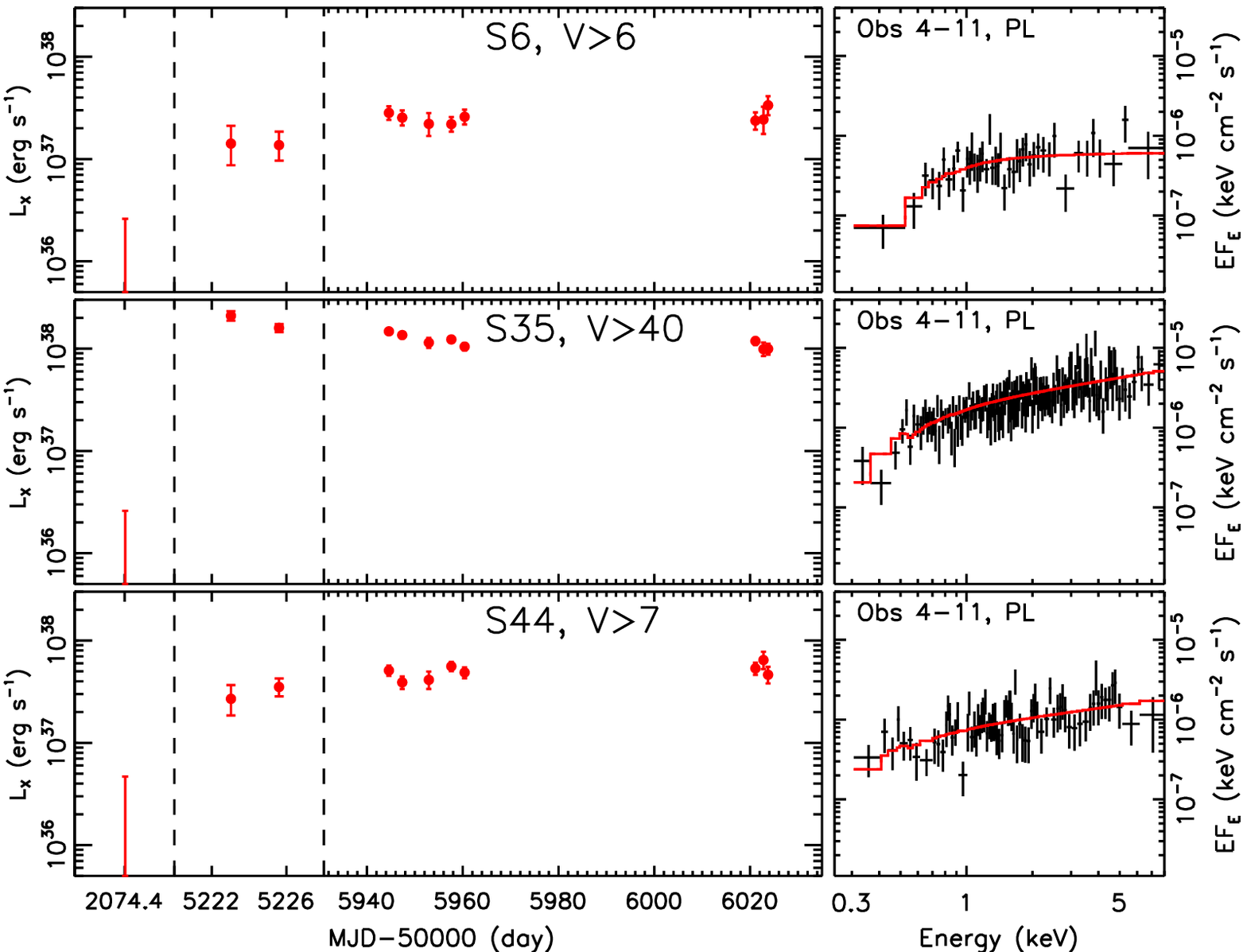}%
    \label{fig:lc_sub_epoch23}
  }
  \caption{(continue)}
  \label{fig:lc_sub}
\end{figure*}

\begin{figure*}
  \centering
  \ContinuedFloat
 \subfloat[A SSS with $kT_\mathrm{BB}=86_{-12}^{+4}$ eV]{%
    \includegraphics[width=0.8\textwidth]{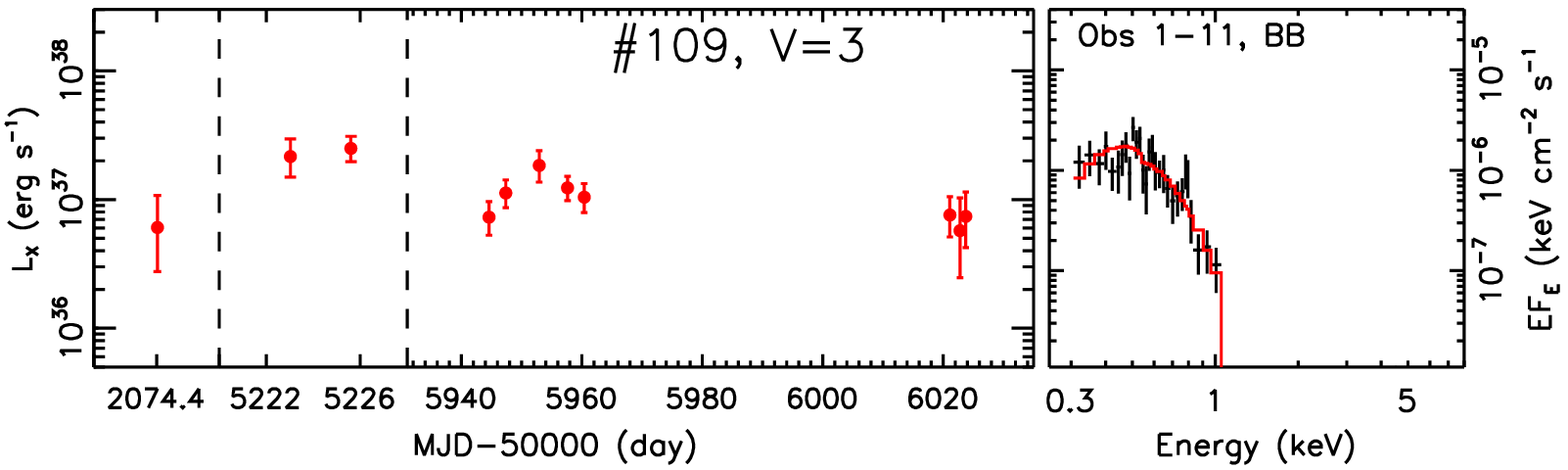}%
    \label{fig:lc_sub_sss}
  }\\
 \subfloat[A very hard X-ray source ($\Gamma_\mathrm{PL}=0.6\pm0.1$) in a
 GC, which is a candidate NS LMXB with a strong magnetic field or a high inclination.]{%
    \includegraphics[width=0.8\textwidth]{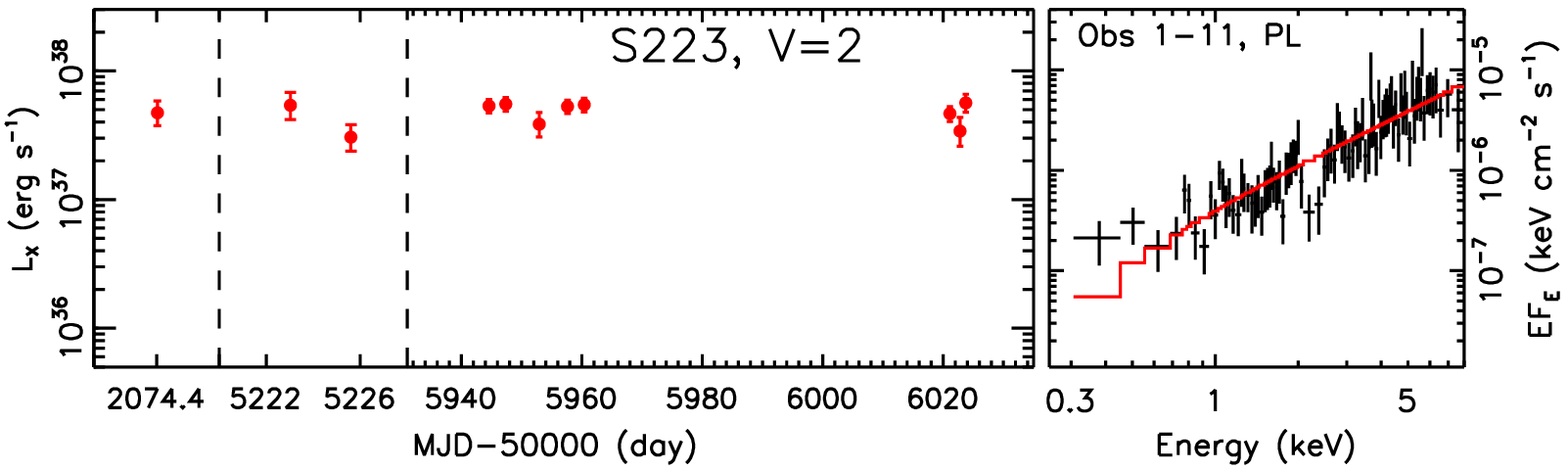}%
    \label{fig:lc_sub_hardsrc}
  }
  \caption{(continue)}
  \label{fig:lc_sub}
\end{figure*}
 
\begin{figure*}
  \centering
  \ContinuedFloat
 \subfloat[Persistent BH X-ray binary candidates. S96 is in a GC while
 others are in the field. S108 has only five detections because it is
 in the CCD gap in other observations. ]{%
    \includegraphics[width=0.8\textwidth]{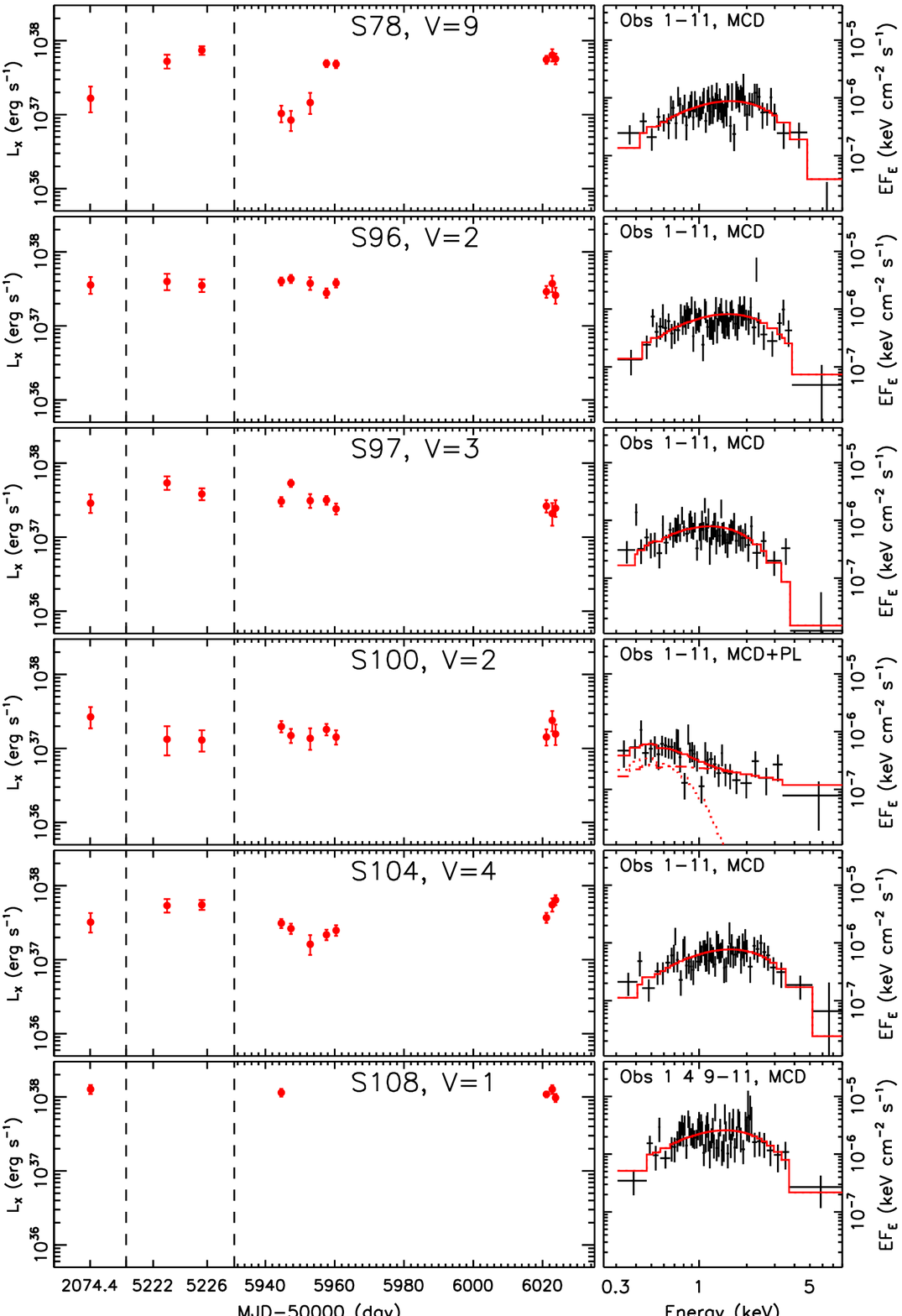}%
    \label{fig:lc_sub_perbh}
  }
  \caption{(continue)}
  \label{fig:lc_sub}
\end{figure*}

\subsection{The Source List and Identification}
\label{sec:srclist}
Figure~\ref{fig:colorimage} shows the false-colored {\it Chandra}
X-ray image of NGC 3115.  We detected 525 unique sources from the
merged and single observations. After eliminating sources below
2$\sigma$ (i.e., the net counts within the 90\% PSF region divided by
the error less than 2), we are left with 490 sources. We found that
the ACIS-S1 chip, which is well outside the galaxy
($>$3$D_\mathrm{25}$), shows some bright streaks, especially at
energies $\lesssim0.7$ keV. We eliminated seven sources detected in
such streaks, as they are most probably spurious. We also eliminated
another source with probably spurious large variability due to being
at the CCD edge. In the end we have 482 sources in total.

Table~\ref{tbl:mscat} gives the various properties of the sources,
such as the flux, the long-term variability, and the maximum
Gregory-Loredo variability index. We also give some source type
information. For the sources within $D_\mathrm{25}$, they are expected
to be dominated by LMXBs and are thus assumed to be such objects,
except those with non-GC optical counterparts, which we identify as
AGNs. The LMXBs identified in this way are expected to be contaminated
by some AGNs, because our optical images are not very deep. Limiting
to the 137 relatively bright sources with
$L_\mathrm{X,max}\gtrsim4\times10^{36}$ erg s$^{-1}$ (i.e.,
  flux $\gtrsim3.6\times10^{-16}$ erg s$^{-1}$ cm$^{-2}$) within
$D_\mathrm{25}$, which are the main targets of our study of this
paper, we identify 9 AGNs. As we found in Paper I that the CXB
  density in our field is consistent with the average value from
  \citet{genala2008}, to within 20\%, we expect $<$16.8 AGNs.  Thus
  the number of AGNs that we might miss is $\lesssim$8, which is
  negligibly small for our purposes. Sources outside
  $D_\mathrm{25}$ should be dominated by AGNs, as we only expect 1.5
  field LMXBs above $4\times10^{36}$ erg s$^{-1}$ based on the IR
  light in the $K_\mathrm{s}$ band outside $D_\mathrm{25}$ (Paper I).
Therefore, we assumed all sources outside $D_\mathrm{25}$ to be AGNs,
except those coincident with GCs, the supersoft X-ray source (SSS)
S109, three coronally active stars in our Galaxy (they are coincident
with stars and show soft X-ray spectra and possible stellar flares),
two galaxies (i.e., due to hot gas emission in galaxies; they are
coincident with galaxies in the optical images and show soft X-ray
spectra), and the BHC S179 (Sections~\ref{sec:ltv} and
\ref{sec:sppr_field}; it is slightly outside $D_\mathrm{25}$ but is
also identified as a field LMXB, considering its possible transient
nature and soft spectra). In total we have 136 candidate field LMXBs
and 49 candiate GC LMXBs (Table~\ref{tbl:gclmxb}). We also marked the
13 transients and 10 BH X-ray binary candidates (BHCs) in the source
type column. Their identification will be described in the following
sections.

Table~\ref{tbl:mscat_sumobs_cfh} gives the counts, fluxes and hardness
ratios of our sources in various energy bands in the merged
observation (also in the high-state and low-state observations for
transients identified in the next section). The counts and fluxes of
our sources in various energy bands in single observations are given
in Table~\ref{tbl:mscat_indobs_cf}.

\subsection{Long-term Variability}
\label{sec:ltv}

The observations of NGC 3115 by {\it Chandra} span more than a decade,
which is ideal for investigation of the long-term variability of
LMXBs. Figure~\ref{fig:ltvar_maxmin} shows the dependence of the
long-term variability $V_\mathrm{var}$ on the maximum 0.5--7 keV
luminosity $L_\mathrm{X,max}$ for all candidate LMXBs except those (24) within
the central elliptical region with semi-major axis $a=10\arcsec$ and
eccentricity and position angle following the $D_\mathrm{25}$ ellipse
(this region is too crowded and the source extraction is subject to
large systematic errors). We find that except for some transients
(triangles, see below), $V_\mathrm{var}$ generally decreases with $L_{\rm
  X,max}$ for both bright GC and field LMXBs with $L_{\rm
  X,max}\gtrsim2\times10^{37}$ erg~s$^{-1}$. At lower luminosities,
$V_\mathrm{var}$ seems to increase with $L_\mathrm{X,max}$, especially for
field LMXBs. However, this is most probably artificial due to the
detection limit of the observations; most of these sources have the
minimum luminosity less than $2\sigma$ and we calculated their $V_{\rm
  var}$ using the $2\sigma$ upper limit of the minimum
luminosity. Some GCs might contain multiple LMXBs (Paper I), but
considering that we detect variability for all GC LMXBs, such source
blending effects should not be significant. The long-term stability of
the most luminous sources that we see in NGC 3115 is also seen in other
galaxies \citep[e.g.,][]{ir2006}.

To search for transients, i.e., quiescent sources with a single
outburst, we concentrated on the 152 sources that are either within
2$D_\mathrm{25}$ (including those within the central $a=10\arcsec$
eclipse) or coincide with GCs and have at least one detection above
4$\sigma$. If we required transients to have $V_\mathrm{var}\ge5$
(there are 25 such sources), be hardly detected (with fluxes
$<2\sigma$) in all observations in at least one epoch, and have the
long-term light curve consistent with a global outburst (instead of an
irregular, large variation), we are left with 13 candidates.
We note that we did not use a very strict condition on the
  variability to select transients, as some persistent Galactic LMXBs
  are known to vary by factors of $>$10
  \citep[e.g.,][]{hokava2009,malokn2010}. Therefore we cannot rule out
  that some transient candidates that we found might be just highly
  variable persistent sources.  We plot the long-term light curves of
the transient candidates in the left panels in Figure~\ref{fig:lc_sub}
a--d. There are three active in the first epoch
(Figure~\ref{fig:lc_sub_epoch1}), one in the second epoch
(Figure~\ref{fig:lc_sub_epoch2}), six in the third epoch
(Figure~\ref{fig:lc_sub_epoch3}), and three active in both the second
and the third epoch (Figure~\ref{fig:lc_sub_epoch23}). We note that
S36 (Figure~\ref{fig:lc_sub_epoch3}), which was bright in the third
epoch, seems to show some emission in the first two epochs, making it
not formally a good transient candidate, but this might be due to
extended emission near the galactic center. We also note that only S8
and S103 are coincident with GCs.

As most of our observations were made within three months in the third
epoch, only outbursts in this epoch are relatively well
monitored. Among the six transient candidates that were active only in
the third epoch (Figure~\ref{fig:lc_sub_epoch3}), three (S103, S125,
and S198) clearly showed flux evolution during the outburst. We
probably detected the decay of the outburst for S103 and S125. For
S198 we had relatively good coverage of the outburst, including the
rise, the peak and the decay. Moreover, for S198, we fortunately
caught a fast rise ($<6$ days), as commonly seen in X-ray binaries
\citep{chshli1997}. These outbursts over month-long timescales are
typical for transient X-ray binaries. For the other three sources (S8,
S26, and S36), the flux remained fairly steady in all seven
observations in the third epoch, indicating relatively long outbursts
(probably years).
\begin{figure*}[!tb]
\centering
\includegraphics[width=4.6in]{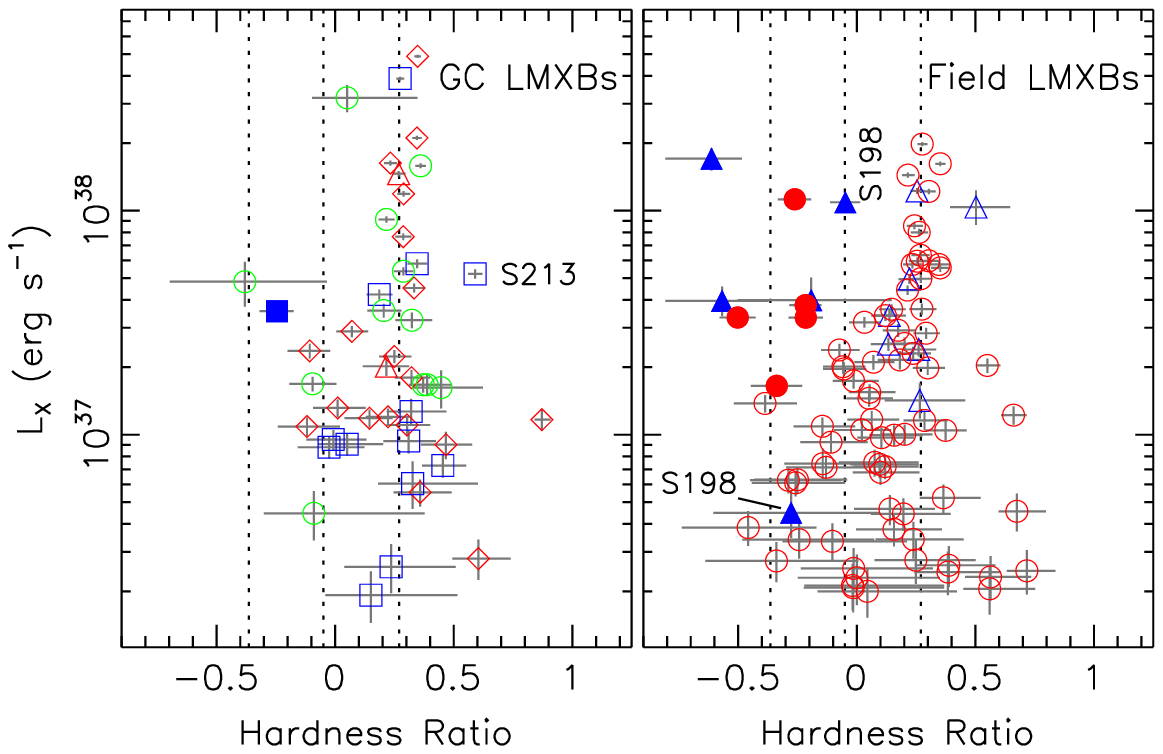}
\caption{The 0.5--7 keV luminosity versus the hardness ratio for
  candidate LMXBs above $4\sigma$ (excluding those in the central
  $a=10\arcsec$ ellipse). The hardness ratio is defined as HR
  $=(H-S)/(H+S)$, where $S$ and $H$ are the energy fluxes in the 0.5--2
  keV and 2--7 keV energy bands, respectively.  The luminosities and
  hardness ratios were calculated from the merged observation, except
  for the 13 transients, for which the merged high-state and low-state
  observations (Figure~\ref{fig:lc_sub}~a--d) were used. The left
  panel is for GC LMXBs, with the blue squares and red diamonds for
  the {\it HST}/ACS blue/metal-poor and red/metal-rich GCs from
  \citet{jestro2014}, respectively, and the green circles for other
  GCs (the Subaru/Suprime-Cam GCs and the extra five {\it HST}/ACS GC
  candidates identified by us; Section~\ref{sec:mulwavcc}). Two
  transients, both in red GCs, are plotted with red triangles. The
  filled square marks the BHC S96. The right panel is for field
  LMXBs. The blue triangles denote the transient candidates. The
  filled circles and triangles denote the BHCs. The error bars in both
  panels are at the $1\sigma$ confidence level. The vertical dotted
  lines from right to left in both panels correspond to hardness
  ratios for an unabsorbed PL with $\Gamma=$1.5, 2.0, and 2.5,
  respectively.}
\label{fig:hr}
\end{figure*}

The three transient candidates that were active in the second and
third epochs had fairly steady fluxes in these two epochs. Because the
second and third epochs together span $\sim$2.2 years, much longer
than typical durations of outbursts seen in X-ray binaries (several
months), these sources were experiencing prolonged outbursts. Such
sources are rare but not unique, as there are several sources known to
have outbursts lasting for years to more than a decade \citep{mcre2006,gamokr2007,sofetu2010,liweba2014}.

The sample spectra of all the transient candidates are shown in the
right panels in Figure~\ref{fig:lc_sub}. Because of large variability
of these sources and in order to increase the statistics, we have
combined observations with similar flux levels, as noted in the
figure. The spectra shown generally represent the high state of the
sources, except S198, for which we also created a low-state spectrum.
While most spectra appear hard and we show the fits with a PL, some
spectra seem soft (e.g., S179), and we show the fits with a MCD. As
will be shown in the next section, these soft sources are most
probably BHBs in the thermal state. We will discuss the spectral
evolution and the physical implication of S198 separately in
Section~\ref{sec:s198}.

\subsection{X-ray Spectral Properties}
\label{sec:sppr}
\subsubsection{The Hardness Ratios}

Figure~\ref{fig:hr} shows the luminosity versus hardness ratio diagram
for our candidate LMXBs, obtained from the merged observation (or the
merged high-state and low-state observations for transients). We
separate the GC (left panel) and field (right panel) sources. For the
former, we further differentiate different subgroups, i.e., the {\it
  HST}/ACS blue/metal-poor GCs in blue squares, the {\it HST}/ACS
red/metal-rich GCs in red diamonds, and the Subaru/Suprime-Cam GCs and
the extra five {\it HST}/ACS GC candidates identified by us
(Section~\ref{sec:mulwavcc}) in green circles.  We observe no clear
spectral differences between these different groups of GC LMXBs,
agreeing with previous findings \citep[e.g.,][]{kikifa2006}.
For both GC and field populations, we find that other than a few very
soft or very hard outliers, our sources seem to follow a global trend
that the luminous sources ($\ge7\times10^{37}$ erg s$^{-1}$) have hard
spectra consistent with $\Gamma_{\rm PL}\sim1.5$, while the fainter
sources have systematically softer spectra. In the following sections
we will present more detailed source spectral properties based on
simple spectral fits and provide systematic comparison with Galactic
LMXBs, which will allow us to shed more light on the nature of our
sources and the cause of their spectral evolution.

\subsubsection{Spectral fits of field LMXBs}
\label{sec:sppr_field}
The results of simple PL and MCD fits to the merged spectra (for
transients, see Figure~\ref{fig:lc_sub} for the spectra used) of
candidate LMXBs are given in Table~\ref{tbl:plmcdfit}.  The C
statistic that we adopted in the fits does not indicate the fitting
quality, but based on bright spectra that were rebinned to have a
minimum of 20 counts per bin and fitted with the $\chi^2$ statistic,
we found that the fits are mostly acceptable with the reduced $\chi^2$
$<$1.2 and the null hypothesis probability $\gtrsim10\%$. There are a
few relatively bad fits using the MCD model, with the reduced $\chi^2$
of 1.4--1.7 and the null hypothesis probability of
$10^{-2}$--$10^{-4}$. Because we are mostly interested in using the PL
and MCD fits to roughly characterize the spectral shapes, the fitting
quality overall is sufficient for our purposes.

\begin{figure*} 
\centering
\includegraphics[width=6.8in]{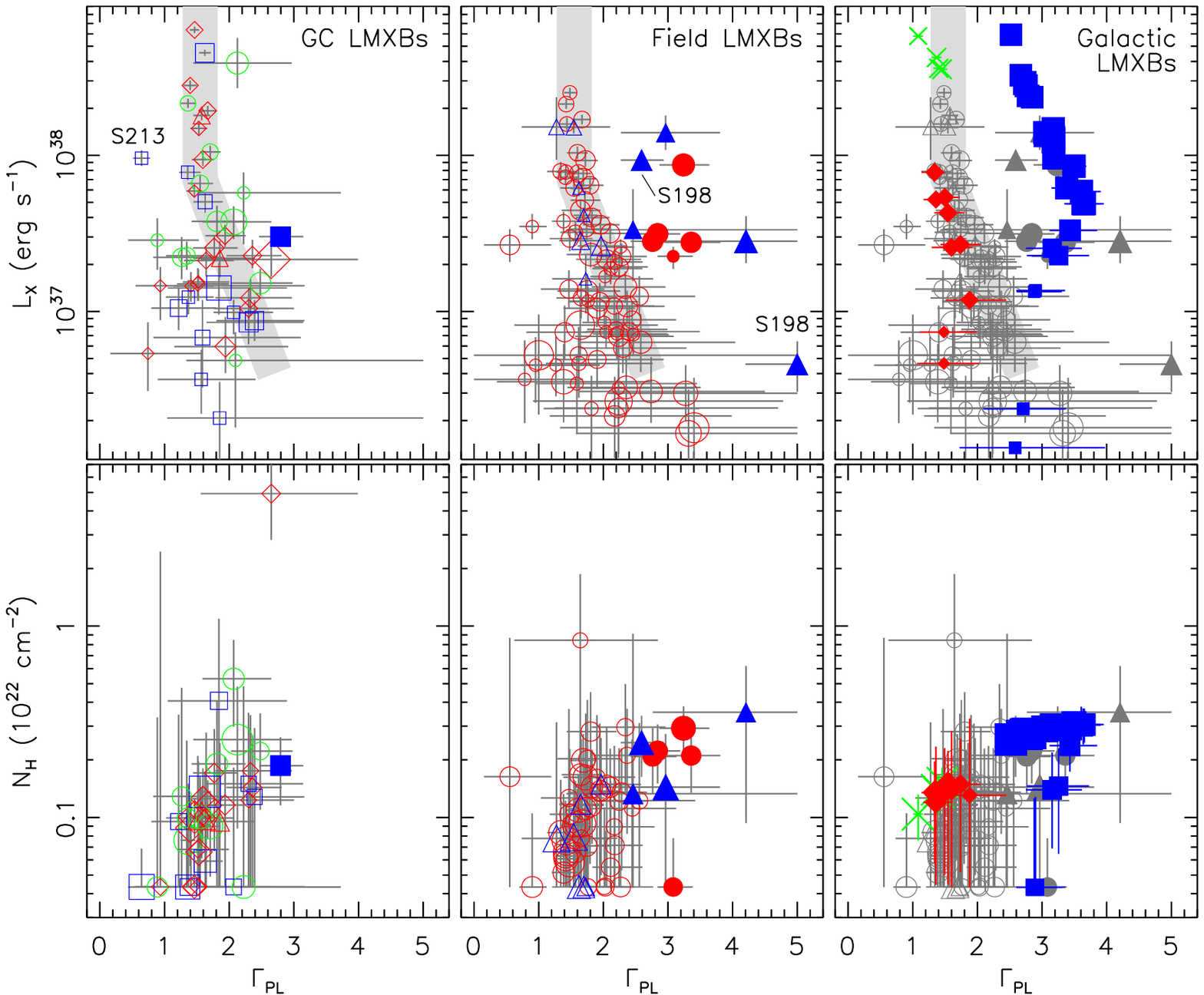}
\caption{The PL fit results of LMXBs excluding those in the central
  $a=10\arcsec$ ellipse, with the top panels plotting the 0.3--8 keV
  luminosity versus the photon index (symbol size proportional to
  column density logarithm) and the bottom panels plotting the column
  density (including the Galactic absorption) versus the photon index
  (symbol size proportional to luminosity logarithm, using data with
  $L_\mathrm{X}>8\times10^{36}$ erg~s$^{-1}$ for clarity). The left
  and middle panels are for GC and field LMXBs, respectively (the
  meanings of the symbols are the same as in Figure~\ref{fig:hr}). The
  right panels plot the PL fits to the atoll source 4U~1705$-$44
  (filled diamonds), the Z source GX~17+2 (crosses) and the BHC
  XTE~J1817$-$330 (filled squares), based on the spectral fits of
  these sources by \citet{lireho2010}, \citet{lireho2012}, and
  \citet{rymist2007}, respectively, and assuming them to be at the
  distance of NGC 3115 with only Galactic absorption ($N_{\rm
    H}=4.32\times10^{20}$ cm$^{-2}$). In these panels, we also plot
  the data shown in the middle panels for field LMXBs, but in a gray
  color. The error bars (sometimes smaller than the symbol size) in
  all panels correspond to the 90\% confidence level. The light gray
  region in the top panels marks the possible NS LMXB soft state
  track, where atolls in the soft state and Z sources reside. }
\label{fig:plindexflux}
\end{figure*}

\begin{figure*} 
\centering
\includegraphics[width=6.8in]{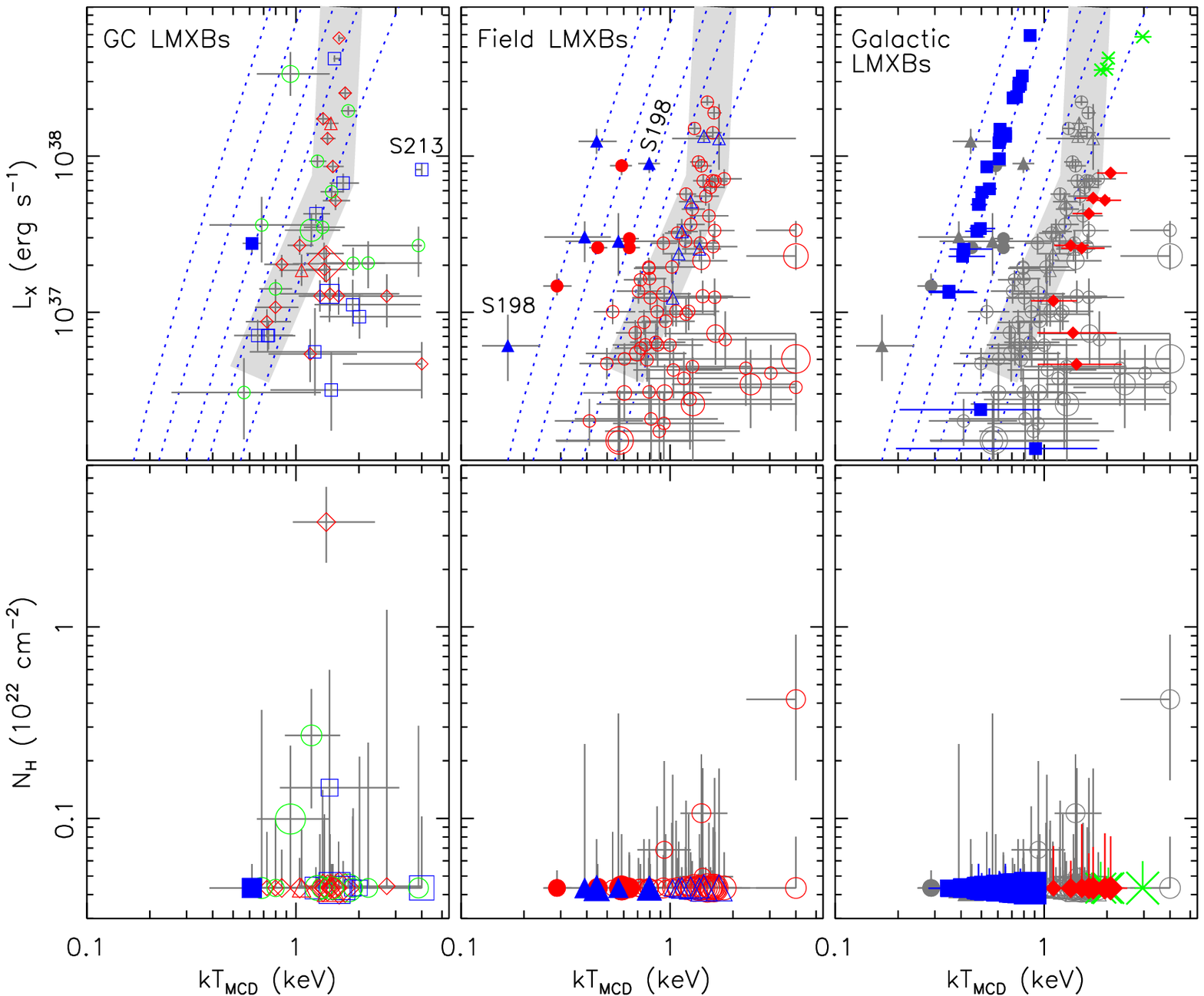}
\caption{Similar to Figure~\ref{fig:plindexflux} but for the MCD fits.
  The dotted lines in the upper panels show the expected dependence of
  the 0.3--8 keV luminosity $L_\mathrm{X}$ on $kT_\mathrm{MCD}$ from the
  standard thermal disk truncated at the ISCO of compact objects of several masses ($M=$ 1.4 \msun, 3 \msun,
  5 \msun, 10 \msun, and 20 \msun\ corresponding to lines from the
  right to the left, respectively; see text for details. The light gray
  region in the top panels marks the possible NS LMXB soft state track, where
  atolls in the soft state and Z sources reside.}
\label{fig:mcdktflux}
\end{figure*}

The PL fits are shown in Figure~\ref{fig:plindexflux}, with the 0.3--8
keV $L_\mathrm{X}$ versus $\Gamma_\mathrm{PL}$ in the top panels and
$N_{\rm H}$ versus $\Gamma_\mathrm{PL}$ in the bottom panels, while
the MCD fits are shown in Figure~\ref{fig:mcdktflux}, with
$L_\mathrm{X}$ versus $kT_\mathrm{MCD}$ in the top panels and
$N_\mathrm{H}$ versus $kT_\mathrm{MCD}$ in the bottom panels. The GC
and field LMXBs are plotted separately, with GCs in the left panels
and field LMXBs in the middle panels. In Figure~\ref{fig:mcdktflux},
some dotted reference lines are included to show the expected
dependence of the 0.3--8 keV $L_\mathrm{X}$ luminosity on
$kT_\mathrm{MCD}$ from the standard thermal disk truncated at the ISCO
of compact objects of several masses ($M=$ 1.4 \msun, 3 \msun, 5
\msun, 10 \msun, and 20 \msun). We used the empirical relation between
the real inner disk radius $r_\mathrm{in}$ and the normalization of
the MCD model $N_\mathrm{MCD}$ \citep{kutama1998,makumi2000}:
\begin{equation}
r_\mathrm{in}=1.19\sqrt{\frac{N_\mathrm{MCD}d_\mathrm{10 kpc}^2}{\cos \theta}}\label{eq:rin}
\end{equation}
where $d_\mathrm{10 kpc}$ is the source distance in units of 10 kpc and
$\theta$ is the inclination angle. The relation takes into account the
spectral hardening effect, with the hardening factor assumed to be
1.7, and the fact that the disk temperature does not peak at the inner
radius. The dotted reference lines in Figure~\ref{fig:mcdktflux}
assume $\theta=60\degr$.

Here we focus on field LMXBs first, with GC LMXBs to be presented in
the next section. The most striking result of the PL fits to field
LMXBs is the strong dependence of the photon index on the luminosity,
as shown in the top middle panel in Figure~\ref{fig:plindexflux}. Most
sources fall on a narrow track (the light gray region), with
$\Gamma_\mathrm{PL}$ decreasing (thus the sources becoming harder)
with increasing $L_\mathrm{X}$ up to $L_\mathrm{X}\sim7\times10^{37}$
erg~s$^{-1}$ and then remaining at a value around 1.5 above this
luminosity. Such a trend has been indicated, though with larger
scatters, in the luminosity versus hardness diagram in
Figure~\ref{fig:hr}. A few very soft outliers lying on the right of
the light gray track can also be seen. Five of them are persistent
sources (filled circles), and the other four are transients (filled
triangles; S198 has two data points, corresponding to its high and low
states). At $L_\mathrm{X}\lesssim2\times10^{37}$ erg~s$^{-1}$, some
hard sources with $\Gamma_\mathrm{PL}\lesssim 1.8$ are also
present. We note that the lack of hard sources with
$\Gamma_\mathrm{PL}\lesssim 1.8$ at $L_\mathrm{X}\lesssim3\times
10^{36}$ erg~s$^{-1}$ is due to selection bias because we only fitted
sources above $4\sigma$ and harder sources tend to have lower
significance levels at a given luminosity. The column density was
inferred to be $\lesssim0.4\times10^{22}$ cm$^{-2}$ for most
sources. It seems to increase with the photon index, which
  could be caused by the use of a (non-physical) PL model to fit
  the spectra that are mainly thermal and are softer at lower
  luminosity (more discussion on the column density is given at the
  end of the section).

The MCD fits give results consistent with the PL fits. Most sources
reside in the light gray region below the $M=3$ \msun\ dotted reference
line (thus corresponding to lower mass) in the top middle panel
plotting $L_\mathrm{X}$ verus $kT_\mathrm{MCD}$ in
Figure~\ref{fig:mcdktflux}, with $kT_\mathrm{MCD}$ increasing with
$L_\mathrm{X}$ below $\sim$$7\times10^{37}$ erg~s$^{-1}$ and then
remaining at around 1.5 keV above this luminosity. The very soft
outliers identified from the PL fits above fall in the region
corresponding to $M\gtrsim5$ \msun. The column density of most sources
from the MCD fits is at the minimum value allowed in the fits, i.e.,
the Galactic value.

To shed light on the possible nature of the sources, in the right
panels in both Figures~\ref{fig:plindexflux} and \ref{fig:mcdktflux},
we plot the expected single-component (PL and MCD) fitting results of
three representative Galactic X-ray binaries, the atoll source
\object{4U 1705-44} (filled red diamonds), the Z source \object{GX
  17+2} (green crosses) and the BHC \object{XTE J1817-330} (filled
blue squares), based on the spectral fits of these sources by
\citet{lireho2010}, \citet{lireho2012}, and \citet{rymist2007},
respectively, and assuming them to be at the distance of NGC 3115 with
absorption at the Galactic value in the direction of NGC 3115 (i.e.,
assuming zero intrinsic absorption). One main reason for us to choose
these sources to compare with our sample is that they had high-quality
broad-band spectra spanning large luminosity ranges and were carefully
modeled in the above studies. The spectra of \object{4U 1705-44}
studied by \citet{lireho2010} used broad-band observations by {\it
  Suzaku} (1.2--40 keV) and {\it BeppoSAX} (1--150 keV) and included
two hard-state spectra and seven soft-state spectra over a large
dynamical range ($\sim$0.04--0.28 $L_\mathrm{Edd}$). For \object{GX
  17+2}, \citet{lireho2012} compiled spectra over the whole Z track in
the HID and used {\it RXTE} data ($\sim3$--40 keV), but for clarity we
only used spectra at four critical positions: the two vertices and the
two ends of the Z track in the HID. We select \object{XTE J1817-330}
studied by \citet{rymist2007} to represent the spectral properties of
BHBs because the outburst in their study was covered by {\it
  Swift}/XRT in the soft X-ray energy band of 0.5--10 keV, close to
the {\it Chandra} bandpass, and spanned a large dynamical range, while
the source is also one of the Galactic BHCs with the lowest
absorption ($N_\mathrm{H}=6\times10^{20}$ cm$^{-2}$). Several spectral
models were applied to these three sources in the above three studies,
but we used the fits of the models that carefully addressed
Comptonization, i.e., the SIMPL(MCD)+BB model, the MCD+BB+nthcomp
model and the MCD+comptt model in \citet{lireho2010},
\citet{lireho2012}, and \citet{rymist2007}, respectively. This is
important because Comptonization with a steep slope can significantly
overestimate the low-energy emission if it is fitted with models like
a simple PL.

To convert the source flux to that at the distance of NGC 3115, we
have adopted the source distance of $d=7.4$ kpc for \object{4U
  1705-44} \citep{hati1995}, $d=8.1$ kpc for \object{GX 17+2}
\citep{lireho2012}, and $d=10$ kpc for \object{XTE J1817-330}
\citep[as assumed in][]{rymist2007}. The distance of \object{4U
  1705-44} was derived from type I X-ray bursts and has a
relatively small uncertainty ($\sim$20\%) depending on the abundance
of the accreted material assumed \citep{gamuha2008}. The distance of
\object{GX 17+2} was also derived based on type I X-ray bursts, but
the bursts in this source exhibited abnormal properties, which could
lead to a relatively large uncertainty ($\sim$50\%) on the distance
derived \citep{lireho2012}. There is no reliable distance estimate for
\object{XTE J1817-330}. The distance of 10 kpc implied a BH of 10
\msun, based on Equation~\ref{eq:rin} \citep{rymist2007}. As long as
the central BH is not too small and the Eddington ratio of the thermal
state in this source is not too low ($\gtrsim1$\%), we expect its
distance to be close to 10 kpc (within a factor of 2). The above
distance uncertainties of \object{GX 17+2} and \object{XTE J1817-330}
are small enough and do not significantly affect our comparison of
these sources with the sources in NGC 3115.

For each spectrum of the representative Galactic X-ray binaries, we
generated 200 simulations and fitted them with PL and MCD models. The
data plotted in Figures~\ref{fig:plindexflux} and \ref{fig:mcdktflux}
are the median of the fitting results, with the error bars
representing the intervals including 90\% of the fits. In the PL and
MCD fits to the simulated 0.3--8 keV spectra of the seven soft-state
observations of \object{4U 1705-44}, we inferred the photon index to
decrease from 1.9 to 1.3 (the top right panel in
Figure~\ref{fig:plindexflux}) or the disk temperature increasing from
1.1 keV to 2.1 keV (the top right panel in Figure~\ref{fig:mcdktflux})
as the 0.3--8 keV luminosity increases from $10^{37}$ erg~s$^{-1}$ to
$7\times10^{37}$ erg~s$^{-1}$. Such trends are very similar to those
traced out by a majority of our field LMXBs in the light gray region
in Figures~\ref{fig:plindexflux} and \ref{fig:mcdktflux} in a similar
luminosity range. \object{GX 17+2} shows similarly hard spectra to the
brightest observation of \object{4U 1705-44} in 0.3--8 keV, with
$\Gamma_\mathrm{PL}\sim1.4$ from the PL fits and $kT_\mathrm{MCD}\sim
2$ keV from the MCD fits. Such properties are similar to those of the
brightest field sources.

Based on the above comparison, we suggest
that the light gray region in Figures~\ref{fig:plindexflux} and
\ref{fig:mcdktflux} marks a NS LMXB soft state track, with sources
below $L_\mathrm{X}\sim7\times10^{37}$ erg~s$^{-1}$ being atolls in
the soft state and brighter sources being Z sources. Then our
observation that most of our sources appear harder at higher
luminosity in 0.3--8 keV can be easily explained because for the atoll
soft state, the temperatures of the disk and boundary layer thermal
emission increase with luminosity at relatively constant emission
areas \citep{lireho2007,lireho2009,lireho2010}. Such trends are not
seen in Z sources because the inner disk and the boundary layer reach
the local Eddington limit, in which case the change in the accretion
rate tends to lead to change in the emission area (the inner radius
for the disk), instead of the temperature \citep{lireho2009}. We note
that our sources seem systematically slightly softer than the
simulated spectra of Galactic NS LMXBs, and one explanation for this
is the possible presence of soft excess of the real disk spectrum
compared with the simple MCD description (i.e., {\it diskbb} in XSPEC)
used in \citet{lireho2010} and \citet{lireho2012}, which will be
discussed in Section~\ref{sec:doublethermalmethod}.

Although the soft-state observations of \object{4U 1705-44} in
\citet{lireho2010} are above $L_\mathrm{X}\sim10^{37}$ erg~s$^{-1}$,
atolls in the soft state can be fainter. For example, the soft state
of \object{4U 1608-52} has Eddington ratios as low as 0.01
\citep{lireho2007}, four times lower than \object{4U 1705-44} in
\citet{lireho2010}. Therefore, the atoll soft-state explanation could
apply to our relatively soft ($\Gamma_\mathrm{PL}\gtrsim2.0$) faint
sources with $L_\mathrm{X}$ as low as a few $10^{36}$ erg~s$^{-1}$.

As mentioned above, we also have nine very soft outliers (filled
circles and triangles in the middle panels in
Figures~\ref{fig:plindexflux} and \ref{fig:mcdktflux}). They are much
softer than NS LMXBs at corresponding luminosities, but are similar to
the BHC \object{XTE J1817-330} in the thermal state. Therefore, we
identified them as BHCs. Their long-term luminosity curves and sample
spectra are shown in Figure~\ref{fig:lc_sub} (those fitted with a MCD
or a MCD+PL model), and the PL and MCD fit results are given in
Table~\ref{tbl:plmcdfit}. The spectra of most of these BHCs can be
described with a MCD, except S100. This source seems to show a hard
tail, and we tried to fit it with a MCD plus a PL, with the photon
index fixed at a value of 2.5. The fitting result is included in
Table~\ref{tbl:spfitextra}, indicating the presence of a PL at the
$6\sigma$ confidence level. S198 is an interesting BHC that will be
described separately in Section~\ref{sec:s198}. It is the only source
with two spectra (one in the high state and the other in the low
state, Section~\ref{sec:s198}, Figure~\ref{fig:lc_sub_epoch3}) in
Figures~\ref{fig:plindexflux} and \ref{fig:mcdktflux}.

The faint hard sources with $L_\mathrm{X}\lesssim2\times10^{37}$
erg~s$^{-1}$ and $\Gamma_\mathrm{PL} \lesssim 1.8$ could be atolls in
the hard state, similar to the two hard-state spectra of \object{4U
  1705-44} (the two filled diamonds below $8\times 10^{36}$
erg~s$^{-1}$ in Figures~\ref{fig:plindexflux} and
\ref{fig:mcdktflux}). We cannot rule out the possibility that there
may be some BHBs in the hard state. For example, simulating the
hard-state spectrum of the BHB \object{GX 339-4} studied by
\citet{mihost2006} and fitting with a PL, as we did above for other
three Galactic X-ray binaries, gave
  $L_\mathrm{X}\sim2\times10^{37}$ erg~s$^{-1}$ and
$\Gamma_\mathrm{PL} \sim 1.6$. However, considering that NS LMXBs are
expected to be more common than BHBs, the possibility being BHBs
should be low.

Our PL fits to simulated spectra of three Galactic X-ray binaries in
the soft/thermal state systematically gave column densities larger
than the value assumed in the simulation (i.e., the Galactic value
toward NGC 3115; see the bottom right panel in
Figure~\ref{fig:plindexflux}). Therefore, the column densities from
the PL fits to our sources are systematically overestimated if our
identification of our sources (i.e., most sources are NS LMXBs in the
soft state with the very soft outliers being BHBs in the thermal
state) is correct. On the contrary, the MCD fits tend to underestimate
the column density, due to the presence of a hard component (weak
Comptonization in the case of BHBs in the thermal state or the
boundary layer emission in the case of NS LMXBs in the soft state)
and/or possible extra soft emission of the real disk spectrum relative
to the MCD description, which will be discussed in
Section~\ref{sec:dis}.

There is one main caveat of the above comparison of our sources with
Galactic sources to be kept in mind.  To have enough statistics, we
have to use the spectra of our sources accumulated over long exposures
($\sim$1.1 Ms) and spanning more than a decade, though most data were
made from the XVP, which still covered nearly three months. To reduce
this problem, we have created spectra for transients carefully by
combining only observations with similar fluxes. For the other
sources, if $L_\mathrm{X, max}\gtrsim2\times10^{37}$ erg s$^{-1}$, most
sources have the long-term variability $V_\mathrm{var}\lesssim3$, which
is not large enough to affect our comparison with Galactic sources
significantly. However, for fainter sources, their long-term
variability is uncertain, and there might be relatively large
systematic errors produced by the dependence of the spectral
properties on the luminosity obtained.

\subsubsection{Spectral fits of GC LMXBs}
\label{sec:sppr_gc}
On the whole, the simple PL and MCD fits indicate that GC LMXBs have
spectral properties very similar to those of field LMXBs (compare the
left panels for GC LMXBs and the right panels for field LMXBs in
Figures~\ref{fig:plindexflux} and \ref{fig:mcdktflux}). In particular,
most GC sources are also in the NS LMXB soft state track (the light
gray region), which exhibits a trend of decreasing
$\Gamma_\mathrm{PL}$ or increasing $kT_\mathrm{MCD}$ with increase in
$L_\mathrm{X}$ below $L_\mathrm{X}\sim7\times10^{37}$ erg~s$^{-1}$ and
relatively constant $\Gamma_\mathrm{PL}$ ($\sim1.5$) or
$kT_\mathrm{MCD}$ ($\sim1.5$ keV) above this luminosity. As discussed
in the previous section for field LMXBs, most of our GC LMXBs below
and above $L_\mathrm{X}\sim7\times10^{37}$ erg~s$^{-1}$ in this track
could be atolls in the soft state and Z sources, respectively. Some
hard sources with $\Gamma<1.8$ below $L_\mathrm{X}\sim3\times10^{37}$
erg~s$^{-1}$ are also present, and they could be NS or BH LMXBs in the
hard state. It seems that GC LMXBs have more hard sources at
luminosities around $10^{37}$ erg~s$^{-1}$ than field LMXBs. For
example, over the range $L_\mathrm{X}=(1$--$2)\times10^{37}$
erg~s$^{-1}$, there are six out of 10 GC LMXBs with $\Gamma<1.6$, but
there is only one out of 14 field LMXBs with $\Gamma<1.6$, from the PL
fits.

There is one persistent source S96 in a blue/metal-poor GC (the filled
square in the left panels in Figures~\ref{fig:plindexflux} and
\ref{fig:mcdktflux}) whose spectrum is significantly softer than those
of other sources at similar luminosity ($L_\mathrm{X}\sim3\times10^{37}$
erg~s$^{-1}$). Its long-term luminosity curve and merged spectrum are
shown in Figure~\ref{fig:lc_sub_perbh}. We identified it as a
BHC. There is another source S223 much harder than other sources at
similar luminosity ($L_\mathrm{X}\sim10^{38}$ erg~s$^{-1}$), which is a
candidate NS LMXB with a high magnetic field or a high inclination and
will be presented separately in Section~\ref{sec:s223}.

\subsubsection{CXB sources}
\begin{figure*} 
\centering
\includegraphics[width=4.6in]{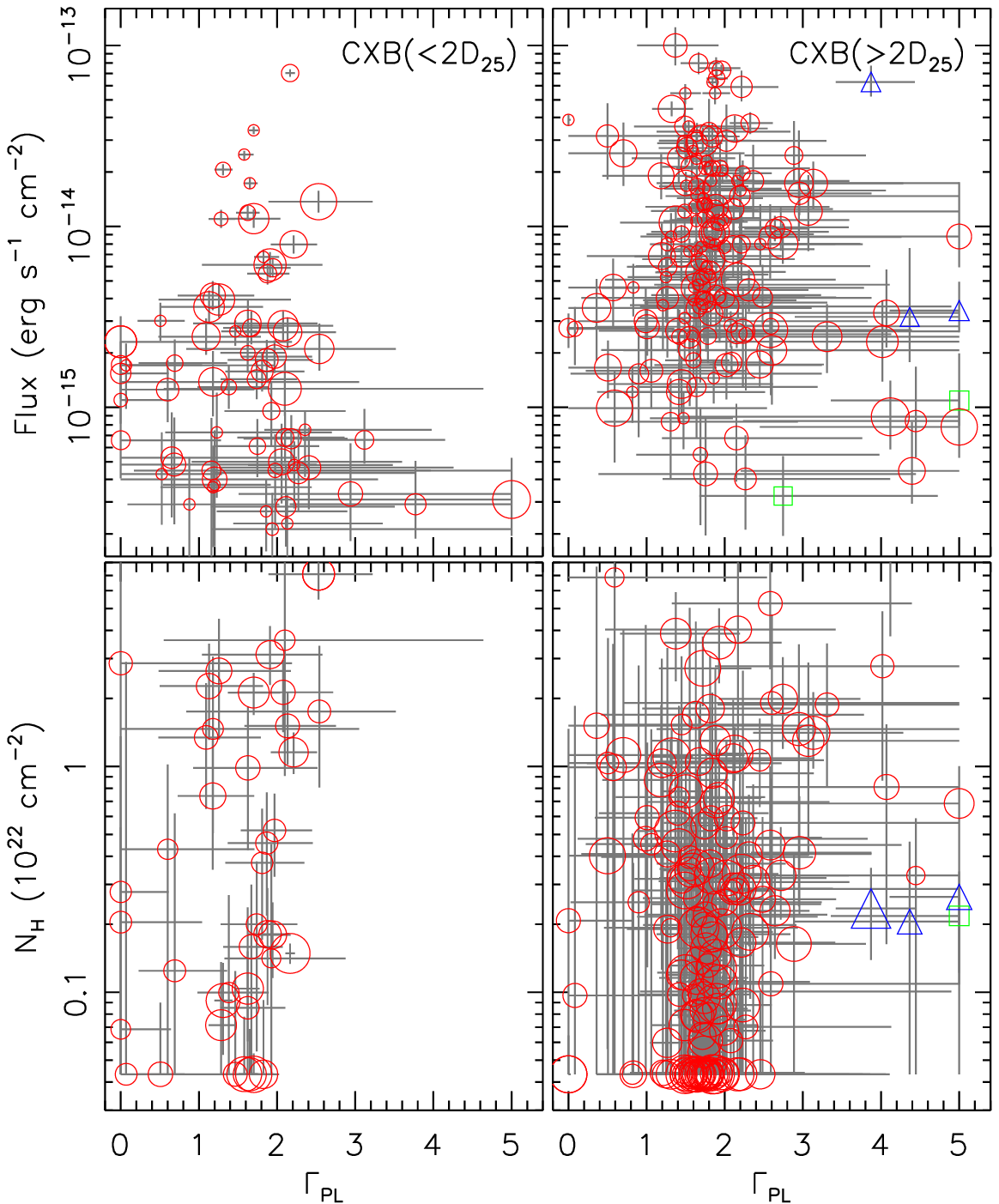}
\caption{Similar to Figure~\ref{fig:plindexflux} but for the PL fits
  to candidate CXB sources.  Sources inside and outside $2D_{25}$ are
  plotted in the left and right panels, respectively. The bottom
  panels for $N_\mathrm{H}$ versus $\Gamma_\mathrm{PL}$ only show sources
  with $L_\mathrm{X}>8\times10^{-16}$ erg~s$^{-1}$~cm$^{-2}$. Three
  candidate coronally active stars (triangles in the right panels) are
  also included. Two faint soft sources (squares in the right panels)
  could be from hot gas emission of optically bright galaxies. }
\label{fig:plindexflux_exga}
\end{figure*}
The results of PL fits to CXB sources are shown in
Figure~\ref{fig:plindexflux_exga}. Sources inside and outside $2D_{\rm
  25}$ are plotted in the left and right panels, respectively. For
both groups of sources, we see no clear dependence of the photon index
on the flux and column density. Using sources outside $2D_\mathrm{25}$
and excluding extreme sources with $\Gamma_\mathrm{PL}>3.5$ and those
with $\Gamma_\mathrm{PL}<0.5$ and limiting to $L_\mathrm{X}>2\times
10^{-15}$ erg~s$^{-1}$ cm$^{-2}$, we obtained a median of $\Gamma_{\rm
  PL}=1.77$ with 68.3\% within 0.38 around it, consistent with the
results from \citet{liweba2012}.
\begin{figure} 
\centering
\includegraphics[width=3.4in]{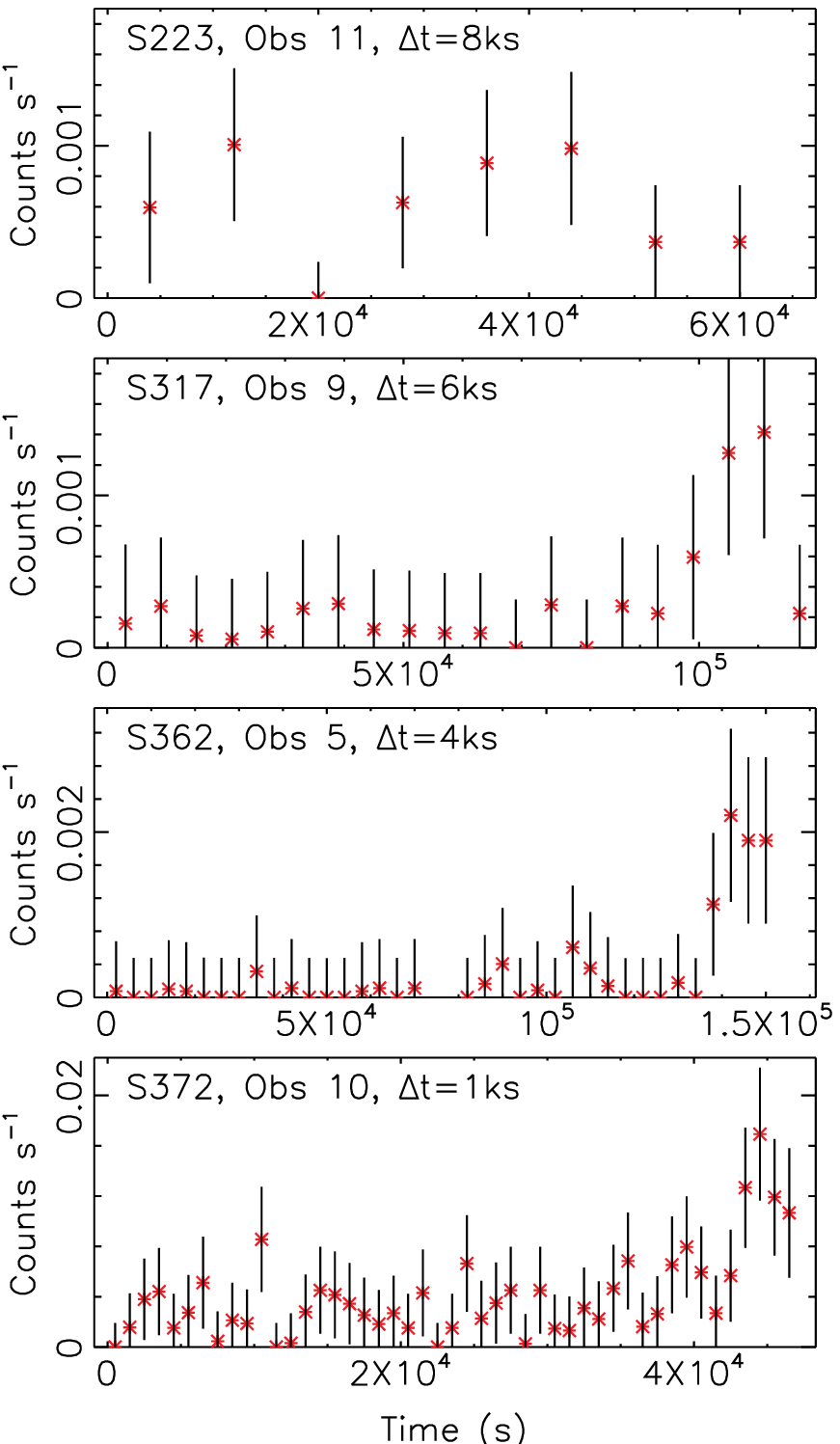}
\caption{The light curves of some sources with interesting
  variability, with Gehrels errors \citep{ge1986} shown. The top panel
  is for S223, which is a candidate NS LMXB with a strong magnetic
  field or a high inclination in a GC, in observation 11, showing
  significant short-term variability. The bottom three panels are for
  three coronally active star candidates in observations showing sign
  of flares. The bin size $\Delta t$ of each light curve is annotated
  in each panel. \label{fig:shorttermlc}}
\end{figure}

In Figure~\ref{fig:plindexflux_exga}, we include three sources
(triangles; S317, S362, and S372) that are candidate coronally active
stars because they all have $\Gamma_\mathrm{PL}>3.5$, coincide with
bright stars in the Subaru/Suprime-Cam imaging, and have possible
stellar flares detected, as shown in the bottom three panels in
Figure~\ref{fig:shorttermlc}. Two faint soft sources ($\Gamma_{\rm
  PL}\gtrsim2.6$; S288 and S290) are plotted with squares. Since they
are coincident with bright extended galaxies in the Subaru/Suprime-Cam
imaging, they might be due to hot gas emission in the galaxies.

\subsection{Special Sources}
\subsubsection{S198: A transient BH LMXB candidate with a prominent
  disk at a very low state}
\label{sec:s198}

As shown in Section~\ref{sec:ltv}, S198 is the only source whose
outburst is covered relatively well by our observations
(Figure~\ref{fig:lc_sub_epoch3}). It also has the most clear spectral
evolution caught among our sources. To accumulate enough statistics,
we combined all the data in the rising (observations 4--5) and decay
(observations 9--11) stages to create a low-state spectrum (the
spectra in these two stages seem to be consistently soft) and the
others (6--8) for a high-state spectrum. When we adopted the PL model,
we obtained $\Gamma_\mathrm{PL}=2.6\pm0.3$ for the high state and
$\Gamma_\mathrm{PL}>4.2$ (the 90\% lower limit) for the low state (the
absorption for the low state was fixed at the value obtained from the
fit to the high state; Table~\ref{tbl:plmcdfit}). Using the MCD model,
we obtained $kT_\mathrm{MCD}=0.78\pm0.07$ keV for the high state and
$kT_\mathrm{MCD}=0.17\pm0.05$ keV for the low state. Therefore, the
spectrum of S198 in the low state in the 2012 outburst is much softer
than in the high state. The 0.3--8 keV $L_\mathrm{X}$ is about 0.12 and
0.008 Eddington luminosity (assuming a BH of mass $M_\mathrm{BH}$=6
\msun\ (see below) and a disk inclination of 60$\degr$) in the high
and low states, respectively.

The disk temperature and luminosity in the high state are typical for
BH X-ray binaries in the thermal state
\citep[Figure~\ref{fig:mcdktflux}; also refer to Figure 16 in
][]{dogiku2007}, in which the thermal disk emission dominates
\citep{remc2006}. Based on Equation~\ref{eq:rin} and the MCD fit in
Table~\ref{tbl:plmcdfit} and assuming a disk inclination of 60$\degr$,
the BH has a mass about 6 \msun.

The low state is more difficult to understand, and to determine the
corresponding spectral state, we compare our source with \object{Swift
  J1753.5-0127} in \citet{rymist2007}.  We first checked for the
possible presence of a PL component in the low state by adding a PL to
the MCD model in the fit to the spectrum.  We fixed
$\Gamma_\mathrm{PL}=1.7$ and $\Gamma_\mathrm{PL}=2.5$, typical values
seen in BHBs, and found that the PL component contributes less than
23\% and 37\% (90\% upper limit) of the 0.3--8 keV unabsorbed flux,
respectively. Thus the MCD still dominates in this energy range in
such fits.  In comparison, \object{Swift J1753.5-0127} has a disk
thermal flux fraction of about 71\%, 62\%, and 56\% in observations
15, 16, and 17 (those with $L_\mathrm{X}\sim1.4\times10^{37}$
erg~s$^{-1}$, $2.5\times10^{37}$ erg~s$^{-1}$, and $3.4\times10^{37}$
erg~s$^{-1}$ in Figures~\ref{fig:plindexflux} and \ref{fig:mcdktflux})
in \citet{rymist2007}, respectively. The disk temperatures are about
0.19 keV, 0.25 keV, and 0.33 keV, and the Eddington ratios are about
0.010, 0.019, and 0.026, assuming a BH mass of 10 \msun\ and a source
distance of 10 kpc, in these three observations, respectively
\citep{rymist2007}. The source was probably in the transitional state
in observations 15 and 16 and in the hard state in observation 17,
based on the broad-band data from the Proportional Counter Array
onboard the \textit{Rossi X-ray Timing Explorer}
\citep{gidopa2008}. While \citet{rymist2007} concluded that the inner
disk radius was consistent with the ISCO in all the three and other
brighter \textit{Swift} observations, \citet{gidopa2008} demonstrated
that the inner disk receded from the ISCO in observations 15--17,
after taking into account the effect of irradiation from the hot
corona on the disk. The inner disk radius in the low state of S198 is
about three times larger than that in the high/thermal state, though
only at the 2.2$\sigma$ level (Table~\ref{tbl:plmcdfit}). Therefore,
based on the disk temperature, the Eddington ratio and the larger
inner disk radius than in the high state, S198 in the low state could
be similar to \object{Swift J1753.5-0127} in observation 17, i.e., in
the hard state. Alternatively and probably more likely, it could be in
the transitional state, considering its more prominent disk than
\object{Swift J1753.5-0127} in observation 17.

\subsubsection{S109: A SSS}
\label{sec:s109}
SSSs have a characteristic temperature $\lesssim$0.1 keV and have been
seen in our Milky Way, the Magellanic Clouds, and nearby galaxies
\citep[for a recent review, see][]{dikopr2010}.  S109 is the only SSS
in our source sample. Its long-term light curve and sum spectrum are
shown in Figure~\ref{fig:lc_sub_sss}. The source seems persistent with
$V_\mathrm{var}=3$. The spectrum can be fitted with a BB with $kT_{\rm
  BB}=86_{-12}^{+4}$ eV, apparent emission radius $R_{\rm
  BB}=3.4_{-0.7}^{+5.3}\times10^3$ km
(Table~\ref{tbl:spfitextra}). The unabsorbed bolometric luminosity is
$(0.8_{-0.2}^{+2.2})\times10^{38}$ erg~s$^{-1}$. SSSs with luminosity
around the Eddington limit for a white dwarf (WD), i.e., $\sim10^{38}$
erg~s$^{-1}$, could be nuclear burning of material accreted by a WD
\citep{gr2000}. Therefore our SSS is probably one such source.

\subsubsection{S223: A very hard luminous X-ray source in a GC}
\label{sec:s223}
As shown in Section~\ref{sec:sppr_gc}, S223 has a spectrum
significantly harder ($\Gamma_\mathrm{PL}=0.6\pm0.1$) than other
sources ($\Gamma_\mathrm{PL}\gtrsim1.4$). The source seems persistent,
with a long-term variability factor of $V_\mathrm{var}=1.8$ at the
2.2$\sigma$ significance level and a mean 0.3--8 keV luminosity of
$0.96\times10^{38}$ erg s$^{-1}$ (Figure~\ref{fig:lc_sub_hardsrc}). We
detected short-term variability in observation 11, with a
Gregory-Loredo variability index of 6 (Figure~\ref{fig:shorttermlc}),
but not in other observations, with Gregory-Loredo variability index
of $\le1$.  The source is coincident with a spectroscopically
confirmed GC, whose $g-z$ color (1.1) is close to the boundary value
of 1.13 used to separate the blue and red GCs (Paper I).  Therefore it
is very unlikely to be a background AGN or a foreground star. It is
probably not composed of multiple sources in the same GC, considering
the detection of both long-term and short-term variability.

There are other X-ray  sources that were found to show very hard spectra and
high luminosities and reside in GCs in nearby galaxies.
\citet{burakr2013} reported a source (their S48) in a GC in Centaurus
A having $\Gamma_\mathrm{PL}\sim0.7$ and the 0.5--10 keV luminosity
$\sim6\times10^{37}$ erg s$^{-1}$.  \citet{trpr2004} reported two
sources (their S22 and S32) in two GCs (Bo91 and Bo158, respectively)
in M31 having $\Gamma_\mathrm{PL}\sim0.6$--0.8 and the 0.3--10 keV
luminosity $\sim$(0.6--1.8)$\times10^{38}$ erg s$^{-1}$. These two
sources showed a spectral cutoff of $\sim$4--8 keV (our source showed
no cutoff below 12 keV at the 90\% confidence level). It is not clear
whether these sources have a common nature. \citet{burakr2013}
  suggested their S48 as a highly magnetized NS as seen in normal
  accreting X-ray pulsars (instead of millisecond X-ray pulsars),
which typically have $\Gamma_\mathrm{PL}\lesssim1$, cut-off energies
around 20 keV, and luminosities $\gtrsim10^{36}$ erg~s$^{-1}$
\citep{whswho1983}. Considering that such objects normally show
coherent pulsations, we searched for them for our source by creating
power density spectra for each observation (refer to
\citet{liweba2014} for the procedure). We found no detection above the
99\% significance level in any observation from the timescale of the
whole observation length up to the Nyquist frequencies (0.62 Hz for
observations 1--3 and 0.64 Hz for observations 4--11). It is possible
that its pulsation period is shorter than the readout frame time or
that the pulsation is not strong enough to be detected with current
data. Most normal accreting X-ray pulsars are high-mass X-ray
binaries, with only a few known to be LMXBs
\citep{bichch1997}. Given the coincidence with a GC, S223 is
more likely to be a LMXB if it is a normal accreting X-ray pulsar.

We note that the hard X-ray source in Bo158 in M31 showed periodic
dips occasionally and thus probably has a high inclination
\citep{trbopr2002,trpr2004}. Therefore we speculate an
  alternative explanation for the hard spectra of the above sources:
  they might have high inclination angles, which suppress the observed
  disk emission and enhance the observed boundary layer emission and
  thus cause the hard X-ray spectra (but the inclination should not be
  too high for the boundary layer to be strongly obscured by the
  disk). Limited by statistics, we cannot determine whether the
short-term variability of our source S223 in observation 11 is due to
dipping.

\section{DISCUSSION}
\label{sec:dis}
\subsection{The NS LMXB Soft State Track and The New Source Identification Scheme}
Except some very soft outliers, which should be strong BHCs in the
thermal state, the majority of our bright sources appear hard in the
{\it Chandra} bandpass (0.3--8 keV). They could be BHBs in the hard
state or NS LMXBs in the soft state. The spectra of NS LMXBs in the
soft state can appear hard in this energy band due to the hot boundary
layer emission. Differentiating between the above two scenarios is
nontrivial based on the narrow-band spectra in hand. The method that
we adopted is to compare the collective spectral properties of our
sources with the spectral evolution of three representative Galactic
X-ray binaries based on simple PL and MCD fits. We found that most of
our sources fall on a narrow track in the $L_\mathrm{X}$ versus
$\Gamma_\mathrm{PL}$ and $L_\mathrm{X}$ versus $kT_\mathrm{MCD}$ plots,
exhibiting harder spectra at higher luminosity below $L_{\rm
  X}\sim7\times10^{37}$ erg~s$^{-1}$ but relatively constant spectral
shape above this luminosity. Such spectral evolution is close to that
expected for NS LMXBs in the soft state in the {\it Chandra}
bandpass. Therefore, we identify the track as the NS LMXB soft state
track, in which sources below $L_\mathrm{X}\sim7\times10^{37}$
erg~s$^{-1}$ are most likely atolls in the soft state and sources
above this luminosity are Z sources.

Our sample of candidate LMXBs in NGC 3115 also includes some hard
sources at low luminosities. Although we believe that they should be
dominated by NS LMXBs, the possibility of some being BHBs (even AGNs)
cannot be ruled out. Therefore our list of BHBs should be
conservative, only including the ten strong BHCs that were identified
based on their very soft spectra. Restricting our search to
(0.046--1.0)$D_{\rm 25}$ and above $10^{37}$ erg~s$^{-1}$, we found
one persistent BHC out of 23 GC LMXBs and nine BHCs (five persistent
and four transient) out of 59 field LMXBs. Therefore, we found a much
larger fraction of BHBs in the field than in GCs, as seen in our
Galaxy. We have identified four BHCs out of 11 field transients. This
is more abundant than the simulations by \citet{frkabe2008} ($<$10\%
for transients above $10^{37}$ erg s$^{-1}$). Our discovery of five
persistent BHCs in the field with $L_\mathrm{X}\gtrsim 10^{37}$ erg is
also not predicted by \citet{frkabe2008}.

X-ray sources in the old populations in other galaxies seem to show
similar spectral properties to those of our sources and can thus be
identified in the same way. \citet{trpr2004} fitted the {\it
  XMM-Newton} (using the 0.3--10 keV band) and {\it Chandra} (using
the 0.5--7 keV band) spectra of 31 bright GC LMXBs in M31 with an
absorbed PL. Their sources had the 0.3--10 keV luminosity $L_\mathrm{X}$
in the range of $\sim10^{36}$ erg s$^{-1}$ to $10^{39}$ erg
s$^{-1}$. In the $L_\mathrm{X}$ versus $\Gamma_\mathrm{PL}$ plot (their
Figure 6), we can see that most of their sources reside in the NS LMXB
soft state track and thus should be atolls in the soft state and Z
sources following our identification scheme, except some hard sources
at low luminosities, which could be NS (most likely) or BH LMXBs in
the hard state. \citet{trpr2004} also argued that most of their
sources, almost all being persistent, should be NS LMXBs, based on the
similarity of their spectral properties and long-term variability to
those of luminous persistent X-ray binaries in our Galaxy (they are
mostly NS LMXBs). However, in their fits to the three brightest
($L_\mathrm{X}>10^{38}$ erg s$^{-1}$) sources with the MCD+BB model, two
have $kT_\mathrm{MCD}\lesssim1$ keV, which seems too low for Z sources,
as will be discussed below.

In the study of X-ray binaries in Centaurus A observed by {\it
  Chandra}, \citet{burakr2013} also obtained the $L_\mathrm{X}$ versus
$\Gamma_\mathrm{PL}$ plot and the $L_\mathrm{X}$ versus $kT_\mathrm{MCD}$ plot
(their Figures 3 and 2, respectively), but for only some bright
sources (0.5--10 keV luminosity $L_\mathrm{X}$ in the range of
$\sim10^{37}$ erg s$^{-1}$ to $4\times10^{38}$ erg s$^{-1}$). They
tried to determine whether the source spectra were dominated by a MCD
or a PL based on the behavior of the inferred column density. Their
method followed \citet{brfabl2010}, who came up with this method based
on simulations of typical BHB spectra. We prefer our source
identification scheme, which is based on the spectral behavior of both
NS and BH LMXBs. Almost all the 17 sources in the $L_\mathrm{X}$ versus
$\Gamma_\mathrm{PL}$ plot in \citet{burakr2013} that they found to be PL
dominated are in the NS LMXB soft state track. Among the 16 sources in
the $L_\mathrm{X}$ versus $kT_\mathrm{MCD}$ plot in \citet{burakr2013} that
they found to be MCD dominated, 12 are in the NS LMXB soft state
track, while the other 4 have very soft spectra ($kT_\mathrm{MCD}\lesssim
1$ keV) and are probably BHBs in the thermal state. \citet{burakr2013}
generally gave consistent source identifications, mostly based on the
disk temperature from the MCD fits only, and our more detailed
comparison with Galactic NS LMXBs makes the identification of the
source nature and spectral state more convincing.

\subsection{Comparison with The Double Thermal Model Source
  Identification Method}
\label{sec:doublethermalmethod}
Given that \citet{lireho2007,lireho2009,lireho2010,lireho2012}
successfully fitted thousands of NS LMXB soft-state spectra with the
double thermal model (MCD+BB), plus a weak Comptonized component when
necessary, \citet{bagamu2013} applied the double thermal model to 35
sources in M31 showing bright hard spectra and found them to be BHCs
because their best-fitting parameters lie significantly outside the
space occupied by NS LMXBs. In particular, they found that the disk
temperatures of their BHCs from the fits with the double thermal model
are systematically lower ($kT_\mathrm{MCD} \lesssim1.0$ keV) than those
of typical NS LMXB soft-state spectra ($kT_\mathrm{MCD} \gtrsim1.0$ keV),
and the BB fractions of the 2--10 keV flux are also larger ($>$65\%
versus $<$50\%).  \citet{bagamu2014} expanded the work and found 15
more BHCs, i.e., 50 in total. These 50 BHCs include 15 GC sources, 12
of which were included in \citet{trpr2004}. Based on our source
identification scheme, however, these 15 BHCs (two are transients) in
GCs should be NS LMXBs. For the other 35 non-GC sources, most of them
should be NS LMXBs too, except for about six sources with very soft
spectra, which should be BHCs.
\begin{figure} 
\centering
\includegraphics[width=3.4in]{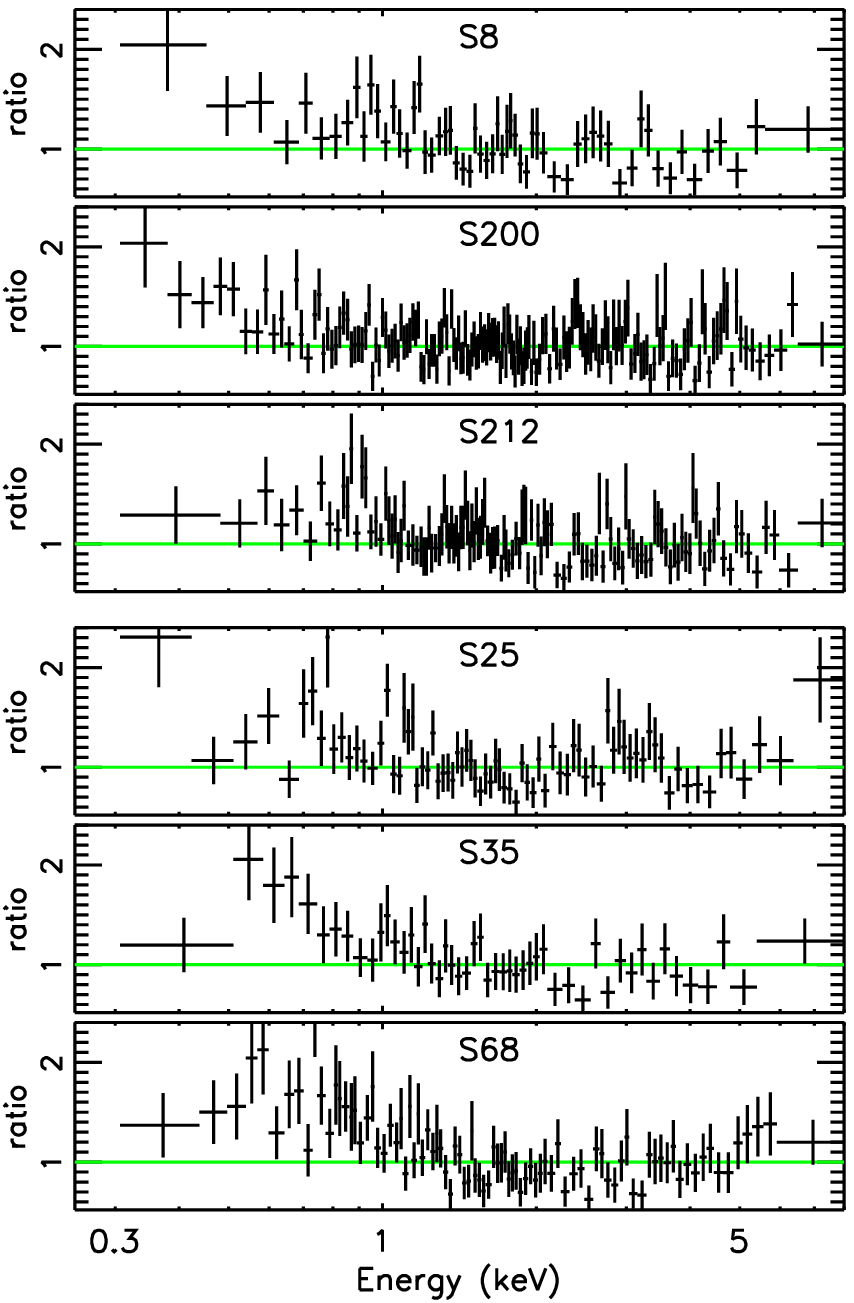}
\caption{The residuals of fits to sample LMXBs with the double thermal
  model with the temperatures of the thermal components fixed at
  values typical of Z sources ($kT_\mathrm{MCD}=1.7$ keV and $kT_{\rm
    BB}=2.5$ keV). }
\label{fig:mcdbbfitratio}
\end{figure}

We also tried to fit the double thermal model to our bright sources
with $L_\mathrm{X}>4\times10^{37}$ erg~s$^{-1}$. There are 33 (14 in GCs
and 19 in the field), excluding the soft BHCs that we have
identified. We found that they all have $kT_\mathrm{MCD}<1$ keV, and 27
of them have the upper error of $kT_\mathrm{MCD}$ at the 90\% confidence
level $<1$ keV. Following \citet{bagamu2013} and \citet{bagamu2014},
this would indicate that almost all bright sources with $L_{\rm
  X}>4\times10^{37}$ erg~s$^{-1}$ in NGC 3115 are BHCs, which seems
very unlikely, considering that BHBs are expected to be much rarer
than NS LMXBs in an early-type galaxy like NGC 3115
\citep{frkabe2008}.

We note that we are using a narrower and softer energy band and have
sources subject to much less absorption than
\citet{lireho2007,lireho2009,lireho2010,lireho2012}. These differences
could cause problems if the double thermal model is directly applied
to our sources. We estimated the possible systematic errors arising
from the use of a narrow energy band. Using the brightest observation
(suz4) of 4U 1705-44 in \citet{lireho2010}, which has $kT_{\rm
  MCD}=1.66$ keV and a BB fraction of 30\% in 2--10 keV, we simulated
200 spectra with Galactic absorption in the direction of NGC 3115 and
then fitted them with the double thermal model with the column density
fixed at the Galactic value. We obtained a median of $kT_{\rm
  MCD}=1.48^{+0.29}_{-0.74}$ keV (the error bar corresponds to the
90\% confidence level in the sense that the upper and lower error bars
each include 45\% of the fits). We also found that 77\% and 46\% of
the simulated spectra have the best-fitting BB fraction in 2--10 keV
larger than 30\% (the BB fraction of suz4) and 50\%, respectively. The
above simulation results indicate that the use of the limited energy
band could systematically infer a lower disk temperature and more BB
contribution in 2--10 keV than expected.

We investigated whether the sources could be fitted well if the
temperatures of the double thermal components are fixed at values
typically seen in NS LMXBs. We concentrate on six bright sources (the
two brightest persistent field LMXBs (S25 and S68), the brightest
transient field LMXB (S35), the two brightest persistent LMXBs in GCs
(S200 in a blue GC and S212 in a red GC) and the brightest transient
GC LMXB (S8); see Table~\ref{tbl:spfitextra}). Because they all have
near or super-Eddington luminosity ($L_\mathrm{X}>10^{38}$ erg~s$^{-1}$,
brighter than the brightest observation (suz4) of 4U 1705-44 in
\citet{lireho2010}), we expect them to be Z sources if they are NS
LMXBs. Then they should have $kT_\mathrm{MCD}$ around 1.7 keV and
$kT_\mathrm{BB}$ around 2.5 keV based on the double thermal model
\citep{lireho2009,lireho2012}. Therefore we also fitted them with the
temperatures of the thermal components fixed at these values. The fit
results are given in Table~\ref{tbl:spfitextra}, and the residuals are
shown in Figure~\ref{fig:mcdbbfitratio}. The fits overall seem fine,
but below around 1 keV, we can see a clear systematic soft excess in
all sources.

One possible explanation of such a soft excess is that the real disk
spectrum is broader than the simple MCD description. A detailed disk
model should carefully calculate radiative transfer through the
vertical structure of the disk and account for the relativistic
smearing. This is implemented in the relativistic disk model {\it
  bhspec}, which indicates that the real disk spectrum should have
excess emission relative to the MCD description by more than 10\% at
low energies \citep{dablhu2005,dahu2006,dadobl2006,kudoda2010}.
Because NSs tend to have a hotter disk, resulting in more of the
low-energy part of the disk spectrum to be observed in the {\it
  Chandra} bandpass than would be the case for BHs, the soft excess of
the real disk spectrum relative to the MCD description is expected to
be more obvious for NSs than for BHs. Such soft excess was not seen by
\citet{lireho2007,lireho2009,lireho2010,lireho2012} because they used
spectra above 1 keV and the sources that they studied had much higher
absorption than the sources studied here.  However, {\it bhspec} has
two limitations, preventing us from applying it to our sources. One is
that it is a table model for the BH accretion disk with the minimum BH
mass $M_\mathrm{BH}$ of 3\msun, not suitable for the accretion disk
around NSs. Besides, the model assumes that the inner disk radius is
at the ISCO, which is not necessarily the case for our luminous
sources, whose inner disk could reach the local Eddington limit and
thus be truncated outside the ISCO \citep{lireho2009}.

Given the above possible soft excess problem, we did not use the
MCD+BB fits to find more BHCs, other than those identified in
Section~\ref{sec:sppr} based on their very soft spectra, a
characteristic that is not seen in NS LMXBs. 

\section{CONCLUSIONS}
\label{sec:con}
We have studied LMXBs detected in NGC 3115 using the Megasecond {\it
  Chandra} XVP observations. Including three previous observations,
the total exposure time is 1.1 Ms. Thus NGC 3115 is one of {\it Chandra}'s
best observed galaxies. In total we have 136 candidate LMXBs in the
field and 49 in GCs detected above $2\sigma$, with $L_\mathrm{X}$ in
the range of $\sim$10$^{36}$ erg~s$^{-1}$ to 10$^{39}$
erg~s$^{-1}$. We calculated the long-term variability for all sources
and identified 13 transient candidates, whose long-term variability
factors are $>5$. Excluding these transients, the sources have
long-term variability overall decreasing with increase in luminosity,
at least at $L_\mathrm{X}\gtrsim2\times10^{37}$ erg~s$^{-1}$.

We carried out simple fits to our sources using single-component
models (a simple PL or a MCD). We found that in the $L_\mathrm{X}$
versus $\Gamma_\mathrm{PL}$ and $L_\mathrm{X}$ versus
$kT_\mathrm{MCD}$ plots a majority of our sources fall on a narrow
track, showing harder spectra at higher luminosity below $L_{\rm
  X}\sim7\times10^{37}$ erg~s$^{-1}$ but relatively constant spectral
shape ($\Gamma_\mathrm{PL}\sim$1.5 or $kT_\mathrm{MCD}\sim1.5$ keV)
above this luminosity. Because our simulations showed that Galactic NS
LMXBs in the soft state show similar spectral evolution in the {\it
  Chandra} bandpass, we identified the track as the NS LMXB soft state
track and suggested sources below $L_\mathrm{X}\sim7\times10^{37}$
erg~s$^{-1}$ as atolls in the soft state and sources above this
luminosity as Z sources. However, the spectra of
our sources seem to show systematic soft excess relative to the double
thermal (MCD+BB) modeling by
\citet{lireho2007,lireho2009,lireho2010,lireho2012} of Galactic NS
LMXBs. One explanation for this is that the real disk spectrum has
excess soft emission relative to the MCD description. This is expected
from detailed simulations involving careful calculation of radiative
transfer through the vertical structure of the disk and accounting for
the relativistic smearing \citep{dablhu2005,dahu2006}.

Ten sources are significantly softer than others at similar
luminosities and are strong BHCs in the thermal state. Five of them are
persistent (one in a blue GC), and the other five are transient.

Some special objects were discovered. S198 is the only transient BHC
whose outburst was covered relatively well by our observations. The
source displayed clear spectral change during the outburst. In the
peak, it is consistent with a BHC of $M_\mathrm{BH}\sim6$ \msun\ in
the thermal state with $kT_\mathrm{MCD}=0.78\pm0.07$ keV and
$L_\mathrm{X}$ at 0.12 Eddington luminosity. The spectrum during the
rise and decay was much softer, with $kT_\mathrm{MCD}=0.17\pm0.05$ keV
and the disk flux fraction $>63\%$ (the 90\% upper limit), and
$L_\mathrm{X}$ was at $\sim$0.008 Eddington luminosity, indicating the
possible presence of a strong thermal disk at a very low Eddington
ratio in a BHB. The source could be in the transitional or hard
state. S109 is a persistent supersoft X-ray source, with $kT_{\rm
  BB}=86^{+4}_{-12}$ eV and $L_\mathrm{BB,
  bol}=0.8^{+2.2}_{-0.2}\times10^{38}$ erg~s$^{-1}$, and can be
explained as due to steady nuclear burning on the surface of a
WD. S223 is a persistent luminous source in a GC with very hard
($\Gamma=0.6\pm0.1$) X-ray spectra source. It is a candidate NS LMXB
with a high magnetic field or a high inclination.

We thank the anonymous referee for the helpful comments. The work is
supported by {\it Chandra} XVP grant GO2-13104X. This material is
based upon work supported in part by the National Science Foundation
under Grants AST-1211995 and AST-1308124. This material is based upon
work supported in part by HST-GO-12759.02-A and HST-GO-12759.12-A. GRS
acknowledges support from an NSERC Discovery Grant.

\clearpage
\tabletypesize{\scriptsize}
\setlength{\tabcolsep}{0.02in}
\begin{deluxetable}{rcccccccccccc}
\tablecaption{The GC LMXBs and Candidates\label{tbl:gclmxb}}
\tablewidth{0in}
\tablehead{
Source&  GC & $\delta_{XO}$ & $g_{\rm ACS}$ & $z_{\rm ACS}$  &$g_{\rm
  SCam}$  & $r_{\rm SCam}$  & $i_{\rm SCam}$ & $R_{\rm h}$ &
Vel\\
& &(arcsec)&(mag)&(mag)&(mag)&(mag)&(mag)&(pc)&(km s$^{-1}$)\\
(1)&(2)&(3)&(4)&(5)&(6) & (7) & (8) &(9) &(10)
}
\startdata
  4 &A4 & 0.06 & $20.087\pm0.002$& $18.786\pm0.002$& $20.021\pm0.003$& $19.268\pm0.004$& $18.968\pm0.003$& $ 2.49$ & 1123\\
  8 &A238 & 0.03 & $23.538\pm0.021$& $21.930\pm0.013$& $23.766\pm0.049$& $22.851\pm0.045$& $22.414\pm0.046$& $ 0.94$ &  \nodata\\
 10 &A41 & 0.19 & $21.296\pm0.007$& $20.250\pm0.010$& $21.085\pm0.107$& $20.413\pm0.122$& $20.274\pm0.116$& $ 1.30$ &  456\\
 11 &A60 & 0.03 & $21.896\pm0.013$& $20.520\pm0.015$& $21.446\pm0.165$& $20.704\pm0.175$& $20.631\pm0.183$& $ 1.36$ &  \nodata\\
 22 &A171 & 0.09 & $22.972\pm0.018$& $21.486\pm0.015$& $23.067\pm0.032$& $22.253\pm0.033$& $21.821\pm0.031$& $ 0.77$ &  \nodata\\
 23 &A15 & 0.03 & $20.782\pm0.003$& $19.473\pm0.003$& $20.672\pm0.005$& $19.937\pm0.005$& $19.644\pm0.005$& $ 1.91$ &  696\\
 24 &A62 & 0.02 & $22.071\pm0.041$& $20.575\pm0.042$& \nodata& \nodata& \nodata& \nodata &  \nodata\\
 70 &A54 & 0.02 & $21.543\pm0.005$& $20.429\pm0.005$& $21.406\pm0.008$& $20.839\pm0.009$& $20.633\pm0.010$& $ 2.23$ &  407\\
 76 &A46 & 0.02 & $21.858\pm0.029$& $20.293\pm0.028$& $20.628\pm0.206$& $19.748\pm0.196$& $19.854\pm0.239$& \nodata &  \nodata\\
 96 &A11 & 0.06 & $20.222\pm0.004$& $19.257\pm0.006$& \nodata& \nodata& \nodata& $ 2.59$ &  \nodata\\
101 &A8 & 0.05 & $20.167\pm0.002$& $19.013\pm0.002$& $20.209\pm0.004$& $19.486\pm0.004$& $19.217\pm0.004$& $ 2.47$ &  \nodata\\
103 &A35 & 0.03 & $21.619\pm0.006$& $20.116\pm0.005$& $21.546\pm0.012$& $20.699\pm0.010$& $20.336\pm0.011$& $ 2.42$ &  821\\
106 &A7 & 0.04 & $20.281\pm0.002$& $18.903\pm0.002$& $20.263\pm0.004$& $19.506\pm0.004$& $19.174\pm0.004$& $ 1.76$ &  697\\
113 &A16 & 0.12 & $20.730\pm0.004$& $19.503\pm0.004$& \nodata& \nodata& \nodata& $ 1.22$ &  \nodata\\
114 &A45 & 0.10 & $21.612\pm0.006$& $20.275\pm0.005$& $21.562\pm0.008$& $20.824\pm0.009$& $20.484\pm0.009$& $ 2.17$ &  \nodata\\
121 &A10 & 0.07 & $20.310\pm0.003$& $19.148\pm0.004$& $20.138\pm0.028$& $19.457\pm0.030$& $19.205\pm0.023$& $ 2.42$ &  \nodata\\
129 &A2 & 0.13 & $19.928\pm0.002$& $18.752\pm0.002$& $19.989\pm0.003$& $19.308\pm0.004$& $18.992\pm0.003$& $ 1.69$ &  \nodata\\
135 &A29 & 0.21 & $20.981\pm0.005$& $20.055\pm0.006$& $21.044\pm0.008$& $20.412\pm0.009$& $20.225\pm0.009$& $ 1.84$ &  \nodata\\
145 &A57 & 0.10 & $21.923\pm0.006$& $20.488\pm0.004$& $22.024\pm0.010$& $21.233\pm0.010$& $20.875\pm0.009$& $ 1.08$ &  798\\
150 &A33 & 0.13 & $21.067\pm0.004$& $20.084\pm0.004$& $21.074\pm0.007$& $20.454\pm0.008$& $20.252\pm0.008$& $ 2.30$ &  949\\
153 &A5 & 0.06 & $20.033\pm0.002$& $18.802\pm0.002$& $20.091\pm0.004$& $19.360\pm0.004$& $19.062\pm0.003$& $ 1.74$ &  682\\
171 &A14 & 0.03 & $20.595\pm0.003$& $19.468\pm0.002$& $20.623\pm0.004$& $19.924\pm0.005$& $19.660\pm0.004$& $ 1.57$ &  \nodata\\
183 &A63 & 0.06 & $21.549\pm0.005$& $20.585\pm0.005$& $21.584\pm0.008$& $20.958\pm0.009$& $20.745\pm0.009$& $ 1.84$ &  441\\
187 &A99 & 0.11 & $22.531\pm0.010$& $20.975\pm0.006$& $22.525\pm0.013$& $21.679\pm0.014$& $21.292\pm0.013$& $ 2.12$ &  609\\
188 &A13 & 0.03 & $20.434\pm0.003$& $19.424\pm0.003$& $20.394\pm0.004$& $19.747\pm0.005$& $19.513\pm0.004$& $ 1.24$ &  426\\
192 &A262 & 0.03 & $23.616\pm0.022$& $22.202\pm0.019$& $23.671\pm0.034$& $22.909\pm0.040$& $22.530\pm0.041$& $ 1.24$ &  \nodata\\
200 &A17 & 0.07 & $20.862\pm0.003$& $19.552\pm0.003$& $20.767\pm0.005$& $20.003\pm0.005$& $19.692\pm0.005$& $ 2.54$ &  439\\
212 &A102 & 0.01 & $21.873\pm0.006$& $21.002\pm0.006$& $21.838\pm0.008$& $21.285\pm0.010$& $21.111\pm0.010$& $ 2.56$ &  \nodata\\
213 &A232 & 0.15 & $22.836\pm0.011$& $21.887\pm0.011$& $22.913\pm0.018$& $22.303\pm0.022$& $22.103\pm0.023$& $ 1.44$ &  885\\
214 &A167 & 0.12 & $22.901\pm0.011$& $21.469\pm0.008$& $22.969\pm0.016$& $22.178\pm0.018$& $21.843\pm0.017$& $ 1.10$ &  750\\
219 &A339 & 0.09 & $24.323\pm0.036$& $23.173\pm0.035$& $24.557\pm0.059$& $23.568\pm0.058$& $23.520\pm0.071$& \nodata &  \nodata\\
223 &A32 & 0.08 & $21.179\pm0.004$& $20.080\pm0.004$& $21.235\pm0.006$& $20.548\pm0.007$& $20.297\pm0.007$& $ 1.44$ &  706\\
226 &A24 & 0.09 & $20.895\pm0.003$& $19.952\pm0.003$& $20.913\pm0.005$& $20.307\pm0.006$& $20.113\pm0.006$& $ 2.37$ &  809\\
229 &A297 & 0.04 & $23.896\pm0.026$& $22.597\pm0.023$& $23.895\pm0.035$& $23.458\pm0.052$& $22.999\pm0.046$& \nodata &  \nodata\\
285 &A249 & 0.37 & $22.933\pm0.013$& $22.051\pm0.016$& $22.935\pm0.019$& $22.393\pm0.024$& $22.185\pm0.025$& $ 2.54$ &  \nodata\\
299 &A18 & 0.15 & $20.981\pm0.004$& $19.641\pm0.003$& $20.877\pm0.005$& $20.109\pm0.005$& $19.805\pm0.005$& $ 2.37$ &  881\\
358 &A192 & 0.97 & $22.516\pm0.008$& $21.628\pm0.008$&$22.528\pm0.012$& $22.044\pm0.016$& $21.797\pm0.015$& $ 4.78$ &  618\\
\hline
199 &S624 & 0.12 & \nodata& \nodata& $22.176\pm0.010$& $21.403\pm0.010$& $21.051\pm0.009$& \nodata&  \nodata\\
322 &S570 & 0.05 & \nodata& \nodata& $23.990\pm0.031$& $23.158\pm0.032$& $22.797\pm0.029$& \nodata&  \nodata\\
325 &S454 & 0.64 & \nodata& \nodata& $22.852\pm0.015$& $22.257\pm0.018$& $21.991\pm0.016$& \nodata&  \nodata\\
332 &S578 & 0.53 & \nodata& \nodata& $22.729\pm0.014$& $21.953\pm0.015$& $21.597\pm0.013$& \nodata&  \nodata\\
356 &S364 & 0.64 & \nodata& \nodata& $22.867\pm0.015$& $22.105\pm0.016$& $21.814\pm0.015$& \nodata&  782\\
378 &S664 & 0.59 & \nodata& \nodata& $22.220\pm0.010$& $21.631\pm0.012$& $21.379\pm0.011$& \nodata&  \nodata\\
451 &S638 & 0.87 & \nodata& \nodata& $19.568\pm0.003$& $18.802\pm0.003$& $18.498\pm0.003$& \nodata&  \nodata\\
\hline
12 &\nodata& 0.06 & $23.377\pm0.045$& $21.832\pm0.042$& \nodata&\nodata& \nodata & $ 0.25$ &    \nodata\\
 53 &\nodata& 0.02 & $21.751\pm0.062$& $20.378\pm0.069$& \nodata& \nodata& \nodata& $ 1.43$ &  \nodata\\
 79 &\nodata& 0.03 & $21.053\pm0.033$& $19.678\pm0.033$& \nodata& \nodata& \nodata& $ 1.01$ &  \nodata\\
 92 &\nodata& 0.07 & $21.693\pm0.006$& $20.306\pm0.005$& $21.738\pm0.011$& $20.959\pm0.011$& $20.608\pm0.010$& $ 1.66$ &  238
\enddata 
\tablecomments{Columns: (1) master source unique index; (2) GC in
  \citet{jestro2014}; (3) the offset between the GC center and our
  X-ray source; (4) {\it HST}/ACS $g$-band magnitude; (5) {\it
    HST}/ACS $z$-band magnitude; (6) SCam $g$-band magnitude; (7) SCam
  $r$-band magnitude; (8) SCam $i$-band magnitude; (9) half-light
  radius; (10) heliocentric velocity, if available, from the
  \citet{poforo2013} catalog. The top group includes 37 LMXBs
  coincident with {\it HST}/ACS GCs. The middle group includes 7 LMXBs
  coincident with GCs detected/covered only in the Subaru/Suprime-Cam
  images, but not in the {\it HST}/ACS images. The bottom group
  includes 4 GC LMXB candidates whose optical matches were not
  classified as GCs by \citet{jestro2014} but were assumed to be GCs
  by us (there is another similar source, S65, which is not included
  in the table because the photometry is not available due to being
  too close to the bright galaxy center; see text for details). Our GC
  LMXB list is slightly different from that in \citet{jestro2014},
  mainly because of our exclusion of very faint X-ray sources
  ($<$2$\sigma$) and update of calibration in our X-ray analysis. }
\end{deluxetable}

\clearpage
\tabletypesize{\normalsize}
\setlength{\tabcolsep}{0.02in}
\begin{deluxetable}{rcccccccccc}
\tablecaption{The master source catalog\label{tbl:mscat}}
\tablewidth{0in}
\tablehead{
Source&  CXOU Name & PU & $\alpha/R_\mathrm{25}$ & $L_{\rm X,max}$ & S/N &$V_\mathrm{var}$ & G-L$_\mathrm{max}$ &Type\\
(1)&(2)&(3)&(4)&(5)&(6) & (7) & (8) &(9)
}
\startdata
 78&J100515.4-074254&0.05&0.168&7.41e+37&  20.3&   9.2&1 &F,BH\\
 96&J100516.2-074235&0.05&0.230&4.33e+37&  19.5&   1.7&2 &GC,BH\\
 97&J100514.2-074233&0.05&0.279&5.41e+37&  19.0&   2.6&7 &F,BH\\
100&J100517.1-074217&0.07&0.323&2.67e+37&  13.3&   2.1&2 &F,BH\\
104&J100516.5-074207&0.06&0.344&6.40e+37&  19.1&   4.0&2 &F,BH\\
108&J100518.5-074138&0.06&0.518&1.27e+38&  19.3&   1.3&0 &F,BH\\
179&J100510.9-074533&0.18&1.040&4.83e+37&   6.2&  19.8&1 &F,BH,T\\
181&J100510.0-074529&0.15&0.936&1.71e+38&   7.9& 126.9&3 &F,BH,T\\
193&J100508.7-074443&0.28&0.571&3.99e+37&   2.9&  30.1&1 &F,BH,T\\
198&J100506.7-074433&0.08&0.711&1.17e+38&  20.2&  33.8&1 &F,BH,T\\
\hline
  8&J100517.2-074352&0.05&0.913&1.60e+38&  37.2&  54.8&2 &GC,T\\
 25&J100516.7-074317&0.04&0.532&1.88e+38&  42.6&   1.5&2 &F\\
 35&J100515.8-074312&0.04&0.349&2.10e+38&  37.1&  40.4&2 &F,T\\
 68&J100513.7-074300&0.03&0.078&2.28e+38&  46.7&   1.5&0 &F\\
200&J100506.0-074428&0.06&0.808&5.71e+38&  69.2&   1.4&0 &GC\\
212&J100527.3-074316&0.07&2.234&4.85e+38&  57.7&   1.4&2 &GC
\enddata 
\tablecomments{This table is published in its entirety in the
  electronic edition of the Astrophysical Journal. A portion is shown
  here, using 10 BHCs (the top group) and 6 bright candidate NS LMXBs
  (the bottom group), for guidance regarding its form and
  content. Columns: (1) master source unique index; (2) IAU name
  (following the convention of CXOU Jhhmmss.s+/-ddmmss); (3)
  1-$\sigma$ statistical positional uncertainty (in units of arcsec)
  in each coordinate, based on Equation (12) of \citet{kikiwi2007};
  (4) The ratio of the angular offset from the galaxy center to the
  elliptical radius of the $D_\mathrm{25}$ isophotal ellipse in the
  direction from the galaxy center to the source; (5) 0.5--7 keV
  maximum luminosity (in units of erg~s$^{-1}$, assuming a source
  distance of 9.7 Mpc); (6) the signal to
  noise ratio (the 0.5--7 keV net counts divided by the error); (7)
  long-term variability; (8) the maximum Gregory-Loredo short-term
  variability index among different observations; (9) the source type
  (field LMXB (``F''), GC LMXB (``GC''), transient (``T''), BHC
  (``BH''), star, AGN, galaxy (``G''), and SSS).}
\end{deluxetable}

\clearpage
\tabletypesize{\normalsize}
\setlength{\tabcolsep}{0.02in}
\begin{deluxetable}{rccccccccccc}
\tablecaption{The Source Counts, Flux, and Hardness Ratio in the Merged Observation\label{tbl:mscat_sumobs_cfh}}
\tablewidth{0in}
\tablehead{
Source&  Obs & Expo &$C_{\rm b}$ & $C_{\rm b}^{\rm l}$ & $C_{\rm b}^{\rm u}$ &$F_{\rm b}$ & $F_{\rm b}^{\rm l}$ & $F_{\rm b}^{\rm u}$
&${\rm HR}_{\rm 3}$ &${\rm HR}_{\rm 3}^{\rm l}$ &${\rm HR}_{\rm 3}^{\rm u}$ \\
(1)&(2)&(3)&(4)&(5)&(6) & (7) & (8) &(9) & (10) & (11) &(12)
}
\startdata
 78 & 0 & 1131.0 & 418.4&  397.8 & 439.2 &3.36e-15 &3.20e-15 &3.53e-15& $-0.19$ &$-0.24$ &$-0.14$\\
 96 & 0 & 1131.0 & 385.9&  366.1 & 405.9 &3.16e-15 &3.00e-15 &3.32e-15& $-0.23$ &$-0.28$ &$-0.18$\\
 97 & 0 & 1131.0 & 363.2&  344.1 & 382.5 &2.96e-15 &2.81e-15 &3.12e-15& $-0.34$ &$-0.39$ &$-0.29$\\
100 & 0 & 1131.0 & 181.8&  168.1 & 195.6 &1.47e-15 &1.36e-15 &1.58e-15& $-0.53$ &$-0.60$ &$-0.47$\\
104 & 0 & 1131.0 & 366.1&  346.9 & 385.5 &2.96e-15 &2.81e-15 &3.12e-15& $-0.19$ &$-0.24$ &$-0.13$\\
108 & 0 &  443.2 & 367.1&  348.0 & 386.3 &9.98e-15 &9.46e-15 &1.05e-14& $-0.27$ &$-0.32$ &$-0.22$\\
179 & 0 & 1131.0 &  50.1&   42.0 &  58.2 &4.40e-16 &3.70e-16 &5.11e-16& $-0.45$ &$-0.59$ &$-0.33$\\
179 & h &  116.3 &  45.2&   38.6 &  52.3 &3.51e-15 &3.00e-15 &4.07e-15& $-0.43$ &$-0.58$ &$-0.32$\\
181 & 0 & 1131.0 &  73.3&   64.0 &  82.7 &6.79e-16 &5.93e-16 &7.66e-16& $-0.18$ &$-0.30$ &$-0.07$\\
181 & h &   35.8 &  74.9&   66.3 &  83.6 &1.51e-14 &1.34e-14 &1.69e-14& $-0.06$ &$-0.18$ &$ 0.05$\\
193 & 0 & 1061.3 &  13.8&    9.0 &  19.2 &1.30e-16 &8.50e-17 &1.80e-16& $-0.05$ &$-0.33$ &$ 0.20$\\
193 & h &   35.8 &  17.5&   13.5 &  22.1 &3.54e-15 &2.74e-15 &4.46e-15& $-0.14$ &$-0.41$ &$ 0.07$\\
198 & 0 & 1131.0 & 413.1&  392.6 & 433.8 &3.82e-15 &3.63e-15 &4.01e-15& $-0.19$ &$-0.25$ &$-0.14$\\
198 & h &  414.9 & 393.2&  373.5 & 413.2 &9.68e-15 &9.19e-15 &1.02e-14& $-0.16$ &$-0.22$ &$-0.11$\\
198 & l &  564.0 &  22.5&   17.4 &  28.2 &3.98e-16 &3.08e-16 &5.00e-16& $-0.77$ &$-0.94$ &$-0.68$\\
\hline
  8 & 0 & 1131.0 &1370.0& 1333.2 &1407.2 &1.10e-14 &1.07e-14 &1.13e-14& $-0.05$ &$-0.09$ &$-0.02$\\
  8 & h &  978.9 &1370.8& 1334.0 &1407.9 &1.30e-14 &1.26e-14 &1.33e-14& $-0.07$ &$-0.10$ &$-0.03$\\
 25 & 0 & 1131.0 &1789.1& 1747.1 &1831.5 &1.44e-14 &1.40e-14 &1.47e-14& $-0.09$ &$-0.12$ &$-0.06$\\
 35 & 0 & 1131.0 &1356.0& 1319.5 &1393.0 &1.10e-14 &1.07e-14 &1.13e-14& $-0.10$ &$-0.13$ &$-0.06$\\
 35 & h &  978.9 &1144.2& 1110.6 &1178.2 &1.09e-14 &1.06e-14 &1.12e-14& $-0.10$ &$-0.13$ &$-0.06$\\
 68 & 0 & 1131.0 &2186.4& 2139.7 &2233.7 &1.76e-14 &1.72e-14 &1.80e-14& $-0.13$ &$-0.16$ &$-0.11$\\
200 & 0 & 1131.0 &4716.3& 4648.2 &4785.1 &4.34e-14 &4.28e-14 &4.40e-14& $ 0.00$ &$-0.02$ &$ 0.02$\\
212 & 0 & 1130.8 &3280.4& 3223.5 &3337.8 &3.45e-14 &3.39e-14 &3.51e-14& $-0.04$ &$-0.06$ &$-0.01$
\enddata 
\tablecomments{This table is published in its entirety in the
  electronic edition of the Astrophysical Journal. A portion is shown
  here, using 10 BHCs (the top group) and 6 bright candidate NS LMXBs
  (the bottom group), for guidance regarding its form and
  content. Columns: (1) master source unique index; (2) observation
  (``0'' refers the combination of all available observations; ``h''
  refers to the combination of the high-state observations for the 13
  transients, as shown in Figure~\ref{fig:lc_sub} a--d; and ``l''
  refers to the combination of the low-state observations for source
  198, as shown in Figure~\ref{fig:lc_sub_epoch3}); (3) exposure time,
  in units of ks; (4-6) the net counts in the broad band (0.5--7.0
  keV) and the lower and upper limits; (7-9) the energy flux in the
  broad band and the lower and upper limits, in units of erg s$^{-1}$
  cm$^{-2}$; and (10-12) the hardness ratio using the fluxes in the
  0.5--2.0 keV and 2.0--7.0 keV energy bands and the lower and upper
  limits.  All limits are at the 68\% confidence level.}
\end{deluxetable}

\clearpage
\tabletypesize{\normalsize}
\setlength{\tabcolsep}{0.02in}
\begin{deluxetable}{rcccccccccc}
\tablecaption{The Source Counts and Flux in Individual Observations\label{tbl:mscat_indobs_cf}}
\tablewidth{0in}
\tablehead{
Source&  ObsID & $C_{\rm b}$ & $C_{\rm b}^{\rm l}$ & $C_{\rm b}^{\rm u}$ &$F_{\rm b}$ & $F_{\rm b}^{\rm l}$ & $F_{\rm b}^{\rm u}$\\
(1)&(2)&(3)&(4)&(5)&(6) & (7) & (8)
}
\startdata
 78 & 13820 &  86.9&   77.6 &  96.4 &4.37e$-$15 &3.90e$-$15 &4.84e$-$15\\
 96 & 13820 &  49.0&   42.2 &  56.5 &2.47e$-$15 &2.13e$-$15 &2.85e$-$15\\
 97 & 13820 &  56.0&   48.5 &  63.5 &2.82e$-$15 &2.44e$-$15 &3.20e$-$15\\
100 & 13820 &  32.1&   26.6 &  38.1 &1.60e$-$15 &1.33e$-$15 &1.91e$-$15\\
104 & 13820 &  38.6&   32.5 &  45.2 &1.93e$-$15 &1.63e$-$15 &2.26e$-$15\\
179 & 13820 &   0.7&    0.0 &   2.8 &3.78e$-$17 &0 &1.46e$-$16\\
181 & 13820 &   0.0&    0.0 &   1.1 &0 &0 &5.97e$-$17\\
193 & 13820 &   0.0&    0.0 &   1.1 &0 &0 &5.87e$-$17\\
198 & 13820 & 174.9&  161.7 & 188.2 &1.04e$-$14 &9.60e$-$15 &1.12e$-$14\\
\hline
  8 & 13820 & 237.2&  221.9 & 252.7 &1.19e$-$14 &1.12e$-$14 &1.27e$-$14\\
 25 & 13820 & 294.4&  277.4 & 311.7 &1.48e$-$14 &1.39e$-$14 &1.56e$-$14\\
 35 & 13820 & 200.9&  186.8 & 215.1 &1.09e$-$14 &1.01e$-$14 &1.17e$-$14\\
 68 & 13820 & 366.4&  347.2 & 385.7 &1.84e$-$14 &1.75e$-$14 &1.94e$-$14\\
200 & 13820 & 707.1&  680.7 & 733.7 &4.23e$-$14 &4.08e$-$14 &4.39e$-$14\\
212 & 13820 & 461.3&  439.9 & 482.8 &3.17e$-$14 &3.02e$-$14 &3.31e$-$14
\enddata 
\tablecomments{This table is published in its entirety in the
  electronic edition of the Astrophysical Journal. A portion is shown
  here, using 10 BHCs (the top group) and 6 bright candidate NS LMXBs
  (the bottom group) in observation 13820, for guidance regarding its
  form and content. Columns: (1) master source unique index; (2)
  observation ID; (3-5) the net counts in the broad band (0.5--7.0 keV)
  and the lower and upper limits; and (6-8) the energy flux in the
  broad band and the lower and upper limits, in units of erg s$^{-1}$
  cm$^{-2}$.  All limits are at the 68\% confidence level.}
\end{deluxetable}

\clearpage
\tabletypesize{\normalsize}
\setlength{\tabcolsep}{0.02in}
\begin{deluxetable}{rcccccccccc}
\tablecaption{Spectral fit results \label{tbl:plmcdfit}}
\tablewidth{0in}
\tablehead{
 & &\multicolumn{4}{c}{the PL Model} && \multicolumn{4}{c}{the MCD Model} \\
 \cline{3-6} \cline{8-11}
Source&Obs&  $N_\mathrm{H}$ & $\Gamma$ &Norm & $L$ & &  $N_\mathrm{H}$ &
$kT$  &$R$ & $L$ \\
(1)&(2)&(3)&(4)&(5)&(6) &&(7) & (8) &(9)&(10)
}
\startdata
78&0&$18.0^{+ 8.4}_{-7.2}$  & $2.84^{+0.35}_{-0.32}$  & $ 1.31^{+  0.39}_{-0.28}$  & $  3.1^{+  0.5}_{ -0.3}$ & &$ 0.0^{+ 2.1}$  & $0.64^{+0.07}_{-0.06}$  & $39.09^{+ 8.86}_{-7.64}$  & $  3.0^{+  0.3}_{ -0.3}$ \\
96 &0& $14.4^{+ 7.6}_{-7.1}$  & $2.79^{+0.35}_{-0.33}$  & $ 1.12^{+ 0.32}_{-0.24}$  & $  3.0^{+  0.4}_{ -0.4}$ & & $ 0.0^{+ 1.4}$  & $0.62^{+0.07}_{-0.06}$  & $40.65^{+ 9.85}_{-8.07}$  & $  2.8^{+  0.3}_{ -0.4}$ \\
97&0&$16.7^{+ 9.0}_{-7.6}$  & $3.36^{+0.45}_{-0.41}$  & $ 1.25^{+ 0.42}_{-0.30}$  & $  2.8^{+  0.4}_{ -0.4}$ & & $ 0.0^{+ 1.4}$  & $0.45^{+0.05}_{-0.04}$  & $76.14^{+18.28}_{-15.39}$  & $  2.6^{+  0.3}_{ -0.2}$ \\
100&0&$ 0.0^{+ 3.5}$  & $3.08^{+0.30}_{-0.24}$  & $ 0.38^{+ 0.06}_{-0.04}$  & $  2.3^{+  0.3}_{ -0.4}$ & & $ 0.0^{+ 1.0}$  & $0.29^{+0.05}_{-0.04}$  & $152.16^{+60.41}_{-49.81}$  & $  1.5^{+  0.3}_{ -0.2}$ \\
104&0&$16.5^{+ 8.5}_{-7.5}$  & $2.76^{+0.37}_{-0.34}$  & $ 1.09^{+ 0.34}_{-0.25}$  & $  2.8^{+  0.4}_{ -0.4}$ & & $ 0.0^{+ 2.3}$  & $0.64^{+0.08}_{-0.07}$  & $36.70^{+ 9.34}_{-7.36}$  & $  2.6^{+  0.2}_{ -0.2}$ \\
108&0&$24.9^{+ 8.8}_{-8.2}$  & $3.24^{+0.40}_{-0.37}$  & $ 4.95^{+ 1.59}_{-1.14}$  & $  8.7^{+  1.1}_{ -0.9}$ & & $ 0.0^{+ 3.9}$  & $0.59^{+0.07}_{-0.07}$  & $79.87^{+24.27}_{-13.20}$  & $  8.7^{+  0.8}_{ -1.1}$ \\
179&h&$31.1^{+26.5}_{-26.1}$  & $4.21^{+0.79}_{-1.44}$  & $ 2.26^{+ 2.77}_{-1.33}$  & $  2.8^{+  1.3}_{ -0.8}$ & & $ 0.0^{+20.1}$  & $0.39^{+0.13}_{-0.14}$  & $111.52^{+279.70}_{-51.07}$  & $  3.0^{+  0.8}_{ -0.9}$ \\
181&h&$10.1^{+13.0}_{-10.1}$  & $2.96^{+0.84}_{-0.69}$  & $ 4.58^{+ 2.88}_{-1.61}$  & $ 14.0^{+  4.0}_{ -3.1}$ & & $ 0.0^{+ 3.5}$  & $0.45^{+0.11}_{-0.08}$  & $171.01^{+92.25}_{-59.61}$  & $ 12.4^{+  2.5}_{ -2.1}$ \\
193&h&$ 9.0^{+77.7}_{-9.0}$  & $2.46^{+2.54}_{-1.03}$  & $ 0.94^{+ 8.05}_{-0.49}$  & $  3.3^{+  2.7}_{ -1.8}$ & & $ 0.0^{+31.0}$  & $0.57^{+0.43}_{-0.25}$  & $49.15^{+201.12}_{-31.44}$  & $  2.8^{+  1.5}_{ -1.1}$ \\
198&h&$20.3^{+ 9.4}_{-8.3}$  & $2.59^{+0.34}_{-0.32}$  & $ 3.63^{+ 1.12}_{-0.84}$  & $  9.3^{+  1.4}_{ -1.2}$ & & $ 0.0^{+ 3.5}$  & $0.80^{+0.10}_{-0.09}$  & $43.19^{+11.46}_{-8.68}$  & $  9.0^{+  0.9}_{ -1.2}$ \\
198 &l& $20.3$(f) & $5.00_{-0.80}$  & $ 0.18^{+ 0.08}_{-0.06}$  & $  0.5^{+  0.2}_{ -0.2}$ & & $ 0.0$(f) & $0.17^{+0.07}_{-0.04}$  & $345.30^{+481.98}_{-235.01}$  & $  0.6^{+  0.4}_{ -0.2}$ \\
\cline{1-11}
8 &h& $ 5.9^{+ 4.2}_{-4.1}$  & $1.57^{+0.13}_{-0.13}$  & $ 2.57^{+ 0.36}_{-0.32}$  & $ 18.0^{+  1.7}_{ -1.2}$ & & $ 0.0^{+ 0.6}$  & $1.47^{+0.13}_{-0.11}$  & $17.12^{+ 2.52}_{-2.28}$  & $ 16.2^{+  1.6}_{ -1.4}$ \\
25 &0& $ 3.2^{+ 3.5}_{-3.2}$  & $1.43^{+0.11}_{-0.11}$  & $ 2.54^{+ 0.29}_{-0.27}$  & $ 21.3^{+  1.5}_{ -1.4}$ & & $ 0.0^{+ 0.6}$  & $1.62^{+0.14}_{-0.10}$  & $15.31^{+ 2.00}_{-1.92}$  & $ 18.9^{+  1.8}_{ -1.7}$ \\
35&h & $ 3.6^{+ 4.3}_{-3.6}$  & $1.54^{+0.14}_{-0.14}$  & $ 2.03^{+ 0.31}_{-0.27}$  & $ 15.1^{+  1.5}_{ -1.3}$ & & $ 0.0^{+ 0.6}$  & $1.45^{+0.14}_{-0.12}$  & $15.91^{+ 2.70}_{-2.40}$  & $ 13.4^{+  1.3}_{ -1.2}$ \\
68 &0& $ 1.5^{+ 3.0}_{-1.5}$  & $1.48^{+0.10}_{-0.09}$  & $ 3.10^{+ 0.33}_{-0.27}$  & $ 25.2^{+  1.8}_{ -1.7}$ & & $ 0.0^{+ 0.4}$  & $1.51^{+0.11}_{-0.10}$  & $19.08^{+ 2.42}_{-2.32}$  & $ 22.2^{+  2.1}_{ -1.0}$ \\
200&0 & $ 5.7^{+ 2.4}_{-2.2}$  & $1.46^{+0.07}_{-0.07}$  & $ 8.14^{+ 0.59}_{-0.55}$  & $ 63.7^{+  3.0}_{ -2.9}$ & & $ 0.0^{+ 0.4}$  & $1.61^{+0.08}_{-0.02}$  & $27.02^{+ 0.54}_{-0.55}$  & $ 57.0^{+  2.7}_{ -2.6}$ \\
212 &0 & $ 9.3^{+ 3.7}_{-3.6}$  & $1.62^{+0.09}_{-0.09}$  & $ 7.09^{+ 0.74}_{-0.61}$  & $ 45.4^{+  2.1}_{ -2.0}$ & & $ 0.0^{+ 0.6}$  & $1.53^{+0.09}_{-0.02}$  & $25.54^{+ 0.51}_{-0.52}$  & $ 42.1^{+  2.0}_{ -1.9}$ 
\enddata 
\tablecomments{This table is published in its entirety in the
  electronic edition of the Astrophysical Journal. A portion is shown
  here, using 10 BHCs (the top group) and 6 bright candidate NS LMXBs
  (the bottom group), for guidance regarding its form and
  content. Columns: (1) master source unique index; (2) the observation
  (``0'' refers the combination of all available observations; ``h''
  refers to the combination of the high-state observations for the 13
  transients, as shown in Figure~\ref{fig:lc_sub}(a)--(d); and ``l''
  refers to the combination of the low-state observations for source
  198, as shown in Figure~\ref{fig:lc_sub}(c)); (3)--(6) the intrinsic
  column density (in units of $10^{20}$ cm$^{-2}$, constrained to be
  $\le 10^{23}$ cm$^{-2}$), photon index (constrained to be $<5.0$),
  normalization (in units of $10^{-6}$ photons keV$^{-1}$ cm$^{-2}$
  s$^{-1}$ at 1 keV), and 0.3-8 keV luminosity (in units of $10^{37}$
  erg s$^{-1}$, assuming a source distance of 9.7 Mpc) corrected for
  Galactic absorption from the fit with the PL model; (7)--(10) the
  intrinsic column density, the maximum disk temperature (in units of
  keV, constrained to be $\le 4$ keV), the apparent inner disk radius
  $R_\mathrm{MCD}$ (in units of km) from the normalization
  $N_\mathrm{MCD}\equiv ((R_\mathrm{MCD}/\mathrm{km})/(D/\mathrm{10\,
    kpc}))^2\cos\theta$, where $D$ is the source distance 9.7 Mpc and
  $\theta$ is the disk inclination assumed to be $60\degr$, and 0.3-8
  keV luminosity corrected for Galactic absorption from the fit with
  the MCD model. All fits used the C statistic. All errors are at the
  90\% confidence level. For S198, the fit to the low-state spectrum
  has the column density fixed (marked with ``f'') at the value
  obtained from the fit to the high-state spectrum.}
\end{deluxetable}

\clearpage
\tabletypesize{\tiny}
\setlength{\tabcolsep}{0.02in}
\begin{deluxetable}{r|c|c|c|c|cc}
\tablecaption{Spectral fit results of some special sources \label{tbl:spfitextra}}
\tablewidth{0pt}
\tablehead{Source & model & $N_\mathrm{H}$ &  Other Parameters &$\chi^2_\nu/\nu$ & $L_{\rm
    abs}$ & $L_{\rm
    unabs}$  \\
             &  & (10$^{20}$ cm$^{-2}$)&      &    & \multicolumn{2}{c}{10$^{36}$ erg~s$^{-1}$}\\
(1)&(2)&(3)&(4)&(5) & (6) &(7)
}
\startdata
100 & MCD+PL & $ 0.0^{+ 7.2}$ &$kT_\mathrm{MCD}=0.14^{+0.05}_{-0.04}$  keV, $R_\mathrm{MCD}=660^{+ 1463}_{ -437}$  km, $\Gamma_\mathrm{PL}=2.5(f)$, $N_\mathrm{PL}=(0.26^{+ 0.07}_{-0.07})\times10^{-6}$ &\nodata& $   22^{+    4}_{   -3}$& $   22^{+   15}_{   -3}$\\
\cline{1-7}
109 & BB & $ 0.7^{+10.6}$ &$kT_\mathrm{BB}=0.086^{+0.007}_{-0.012}$  keV, $R_\mathrm{BB}=3376^{+5272}_{-674}$  km  &\nodata& $   37^{+    7}_{  -10}$  & $   41^{+   79}_{   -8}$ \\
\cline{1-7}
8& MCD+BB & $ 0.0^{+ 3.2}$ &$kT_\mathrm{MCD}= 0.67^{+ 0.34}_{-0.20}$  keV, $R_\mathrm{MCD}=   56^{+   46}_{  -27}$  km, $kT_\mathrm{BB}=1.43^{+1.76}_{-0.35}$  keV, $R_\mathrm{BB}= 14.6^{+ 10.6}_{-11.3}$  km &$0.93( 55)$ & $  166^{+   10}_{  -12}$  & $  166^{+   13}_{  -12}$ \\
25& MCD+BB & $ 0.0^{+11.2}$ &$kT_\mathrm{MCD}=0.46^{+ 0.23}_{-0.18}$  keV, $R_\mathrm{MCD}=100^{+ 238}_{ -51}$  km, $kT_\mathrm{BB}=1.11^{+ 0.32}_{-0.16}$  keV, $R_\mathrm{BB}=26.7^{+ 10.4}_{-10.8}$  km &$1.17( 72)$ & $  184^{+   11}_{  -14}$  & $  184^{+   29}_{  -11}$ \\
35& MCD+BB & $ 0.0^{+ 5.8}$ &$kT_\mathrm{MCD}= 0.52^{+ 0.24}_{-0.18}$  keV, $R_\mathrm{MCD}=   77^{+   98}_{  -21}$  km, $kT_\mathrm{BB}=1.22^{+0.59}_{-0.22}$  keV, $R_\mathrm{BB}= 18.4^{+  9.4}_{-10.0}$  km &$0.74( 45)$ & $  137^{+   11}_{  -11}$  & $  137^{+   12}_{  -10}$ \\
68& MCD+BB & $ 1.7^{+ 5.6}$ &$kT_\mathrm{MCD}=0.42^{+ 0.11}_{-0.09}$  keV, $R_\mathrm{MCD}=154^{+ 128}_{ -59}$  km, $kT_\mathrm{BB}=1.24^{+ 0.21}_{-0.15}$  keV, $R_\mathrm{BB}=24.1^{+  6.4}_{ -5.9}$  km &$0.82( 84)$ & $  231^{+   15}_{  -14}$  & $  239^{+   22}_{  -16}$ \\
200& MCD+BB & $ 0.0^{+ 1.0}$ &$kT_\mathrm{MCD}=0.62^{+ 0.35}_{-0.15}$  keV, $R_\mathrm{MCD}=106^{+  60}_{ -54}$  km, $kT_\mathrm{BB}=1.16^{+ 0.38}_{-0.14}$  keV, $R_\mathrm{BB}=41.5^{+ 13.8}_{-21.3}$  km &$0.96(165)$ & $  556^{+   21}_{  -21}$  & $  556^{+   21}_{  -21}$ \\
212& MCD+BB & $ 0.0^{+ 5.8}$ &$kT_\mathrm{MCD}=0.56^{+ 0.18}_{-0.15}$  keV, $R_\mathrm{MCD}=113^{+ 100}_{ -44}$  km, $kT_\mathrm{BB}=1.15^{+ 0.21}_{-0.13}$  keV, $R_\mathrm{BB}=36.0^{+ 10.1}_{-11.4}$  km &$0.91(123)$ & $  416^{+   16}_{  -20}$  & $  416^{+   30}_{  -16}$ \\
\cline{1-7}
8& MCD+BB & $ 0.0^{+ 0.6}$ &$kT_\mathrm{MCD}=1.70$(f) keV, $R_\mathrm{MCD}= 13.1^{+   0.3}_{  -0.3}$  km, $kT_\mathrm{BB}=2.50$(f) keV, $R_\mathrm{BB}= 0.0^{+  1.6}$  km &$1.37( 57)$ & $  165^{+    7}_{   -7}$  & $  165^{+    7}_{   -7}$ \\
25& MCD+BB & $ 0.0^{+ 0.6}$ &$kT_\mathrm{MCD}=1.70$(f) keV, $R_\mathrm{MCD}= 13.8^{+   0.3}_{  -0.4}$  km, $kT_\mathrm{BB}=2.50$(f) keV, $R_\mathrm{BB}= 0.1^{+  2.6}$  km &$1.38( 75)$ & $  183^{+   11}_{   -7}$  & $  183^{+   11}_{   -7}$ \\
35& MCD+BB & $ 0.0^{+ 0.6}$ &$kT_\mathrm{MCD}=1.70$(f) keV, $R_\mathrm{MCD}= 12.0^{+   0.3}_{  -0.3}$  km, $kT_\mathrm{BB}=2.50$(f) keV, $R_\mathrm{BB}= 0.0^{+  1.5}$  km &$1.32( 47)$ & $  137^{+    6}_{   -6}$  & $  137^{+    6}_{   -6}$ \\
68& MCD+BB & $ 0.0^{+ 0.4}$ &$kT_\mathrm{MCD}=1.70$(f) keV, $R_\mathrm{MCD}= 15.2^{+   0.4}_{  -0.4}$  km, $kT_\mathrm{BB}=2.50$(f) keV, $R_\mathrm{BB}= 0.1^{+  2.2}$  km &$1.73( 86)$ & $  220^{+    9}_{   -8}$  & $  220^{+    9}_{   -8}$ \\
200& MCD+BB & $ 0.0^{+ 0.4}$ &$kT_\mathrm{MCD}=1.70$(f) keV, $R_\mathrm{MCD}= 24.3^{+   0.4}_{  -0.4}$  km, $kT_\mathrm{BB}=2.50$(f) keV, $R_\mathrm{BB}= 0.1^{+  2.4}$  km &$1.05(167)$ & $  565^{+   13}_{  -14}$  & $  565^{+   13}_{  -13}$ \\
212& MCD+BB & $ 0.0^{+ 0.4}$ &$kT_\mathrm{MCD}=1.70$(f) keV, $R_\mathrm{MCD}= 21.0^{+   0.4}_{  -0.4}$  km, $kT_\mathrm{BB}=2.50$(f) keV, $R_\mathrm{BB}= 0.1^{+  1.8}$  km &$1.14(125)$ & $  422^{+   12}_{  -12}$  & $  422^{+   12}_{  -12}$ 
\enddata 
\tablecomments{Columns: (1) master source unique index; (2) the spectral
  model; (3) intrinsic column density (4) other spectral parameters
  ($R_\mathrm{BB}$ is the apparent source radius from the BB
  normalization $N_\mathrm{BB}\equiv ((R_\mathrm{BB}/\mathrm{km})/(D/\mathrm{10
    kpc}))^2$, and refer to Table~\ref{tbl:plmcdfit} for the meanings
  of other parameters); (5) the reduced $\chi^2$ and the degrees of
  freedom, (only for fits to spectra binned to have a minimum of 20
  counts and using the $\chi^2$ statistic), (7)--(8) the absorbed (but
  corrected for Galactic absorption) and unabsorbed 0.3--8 keV
  luminosity (assuming isotropic emission or a disk inclination of
  60$\degr$), respectively. All errors are at the 90\% confidence
  level. }
\end{deluxetable}

\end{document}